\documentclass[lettersize,journal]{IEEEtran}
\usepackage{amsmath,amsfonts,amsthm}
\usepackage{dsfont}
\usepackage{algorithmic}
\usepackage{algorithm}
\usepackage{array}
\usepackage[caption=false,font=normalsize,labelfont=sf,textfont=sf]{subfig}
\usepackage{textcomp}
\usepackage{stfloats}
\usepackage{url}
\usepackage{verbatim}
\usepackage{graphicx}
\usepackage{cite}
\usepackage{hyperref}
\newcommand{\R}{\mathbb{R}}

\newtheorem{lemma}{Lemma}
\newtheorem{theorem}{Theorem}

\begin{document}

\title{Adaptive Probabilistic Forecasting of Electricity (Net-)Load}

\author{Joseph de Vilmarest, Jethro Browell, Matteo Fasiolo, Yannig Goude and Olivier Wintenberger
\thanks{J. de Vilmarest is with Viking Conseil; he was with \'Electricit\'e de France R\&D and Sorbonne Universit\'e during part of the work (e-mail: joseph.de-vilmarest@vikingconseil.fr).}
\thanks{J. Browell (Senior Member, IEEE) is with the University of Glasgow (e-mail:  Jethro.Browell@glasgow.ac.uk).}
\thanks{M. Fasiolo is with the University of Bristol (e-mail: matteo.fasiolo@bristol.ac.uk).}
\thanks{Y. Goude is with \'Electricit\'e de France R\&D and with the Laboratoire de Mathématique d’Orsay, Université Paris-Saclay (e-mail: yannig.goude@edf.fr).}
\thanks{O. Wintenberger is with Sorbonne Universit\'e and with Wolfgang Pauli Institut, c/o Fakultät für Mathematik, Universität Wien (e-mail: olivier.wintenberger@sorbonne-universite.fr).}
}

\markboth{}
{Shell \MakeLowercase{\textit{et al.}}: A Sample Article Using IEEEtran.cls for IEEE Journals}


\maketitle

\begin{abstract}
Electricity load forecasting is a necessary capability for power system operators and electricity market participants. 
The proliferation of local generation, demand response, and electrification of heat and transport are changing the fundamental drivers of electricity load and increasing the complexity of load modelling and forecasting.
We address this challenge in two ways. First, our setting is {\it adaptive}; our models take into account the most recent observations available, yielding a forecasting strategy able to automatically respond to changes in the underlying process.
Second, we consider {\it probabilistic} rather than point forecasting; indeed, uncertainty quantification is required to operate electricity systems efficiently and reliably.
%
%
Our methodology relies on the Kalman filter, previously used successfully for adaptive point load forecasting. The probabilistic forecasts are obtained by quantile regressions on the residuals of the point forecasting model. We achieve adaptive quantile regressions using the online gradient descent; we avoid the choice of the gradient step size considering multiple learning rates and aggregation of experts. We apply the method to two data sets: the regional net-load in Great Britain and the demand of seven large  cities in the United States. Adaptive procedures improve forecast performance substantially in both use cases for both point and probabilistic forecasting.
\end{abstract}


\section{Introduction}
\IEEEPARstart{F}{orecasting} electricity demand is fundamental in the process of maintaining supply-demand balance. This permanent equilibrium is necessary to maintain a reliable supply of electricity and to avoid damaging infrastructure. As electricity cannot be stored on a large scale, forecasts are crucial to informing production planning. This necessity explains why energy forecasting has gathered so much attention from the time series and forecasting community \cite{Hong2020review}. The recent increase in electricity prices in Europe further emphasizes the importance of demand forecast quality.

The literature historically focused on point forecasting. However, the expected value of the load is not sufficient for risk management. Forecasting models cannot be perfect and therefore the production plans cannot match exactly the demand; therefore, grid operators need to call reserves to maintain the equilibrium and traders need to hedge their market positions. It is thus essential to have some information on the distribution around the mean to schedule the reserve needed. In Great Britain, for instance, reserves are tailored to be sufficient in all but 4 hours per year; this corresponds to the estimation of 99.95\% confidence intervals.
The importance of probabilistic forecasts was highlighted in the last two Global Energy Forecasting Competitions \cite{HONG2016896,hong2019global}. Indeed, the objective in both competitions was to forecast certain quantiles, and the submissions were evaluated through the sum of quantile losses.

Demand and supply characteristics are both evolving with time. On the demand side, unexpected events such as  the coronavirus crisis or the recent increase in electricity prices in Europe can affect it significantly and abruptly \cite{IEA2021_covid, IEA2022_outlook}. From a longer-term perspective, the evolution of consumption habits such as those induced by the expected growth in electric vehicle penetration will change the demand patterns. On the production side, the increasing penetration of intermittent electricity production strongly changes the forecasting task. Indeed, controllable production units must be employed to meet the difference between the demand and the intermittent electricity production (mainly wind and solar energy), denoted by {\it net}-load. These changes in the behavior of the net-load can not be captured by classical offline probabilistic methods and motivate the need for adaptive methods, taking into account observations in a streaming fashion to improve the forecasting model.

The recent point forecasting literature has highlighted the improvements yielded by adaptive methods able to learn regime changes. Adaptation is especially crucial to forecast wind and solar power, and it has been successfully applied to online variable selection \cite{messner2019online} and online forecast reconciliation \cite{di2021online}. State-space representations have well captured the recent changes of pattern due to the coronavirus crisis \cite{obst2021adaptive,de2022state}.
While the majority of recent works have focused on point forecasting,  \cite{Alvarez2021} consider adaptive probabilistic load forecasting based on hidden Markov models and is found to perform well for a range of loads; however, this approach relies on Gaussian predictive distributions which leads to relatively poor calibration (quantile bias) on the data we consider here. In \cite{gaillard2016additive, Fasiolo_2020_qgam}, quantile GAMs are proposed, as well as variants of GAMLSS (Location Scale and Shape) in \cite{Gilbert2023}, for probabilistic demand forecasting. These methods relax the Gaussian assumption but are not adaptive. \cite{Wang2022} propose a mixture model for short term probabilistic forecasting of individual load demand at different aggregation level, but the distribution of the data is supposed to be invariant with time. \cite{Arora2022RemodellingSP} propose, similarly to \cite{obst2021adaptive}, a deep neural network probabilistic forecasting method where the features extracted with a multi-layer RNN-LSTM are mapped to a state-space model for online adaptation. Their work is limited to time-varying Gaussian distributions as they use a Gaussian state-space model to generate density forecasts.

We propose in this article to apply the adaptive framework to quantile regression for probabilistic forecasting. There has been scarce work in that direction. We mention the exciting use of aggregation of experts to obtain an online forecaster \cite{gaillard2016additive,berrisch2021crps}. The principle is to combine various predictions (experts) with a weighted average where the weights evolve over time. We exploit the idea of aggregation, but we introduce a more structural adaptation in the sense the experts themselves are adaptive.

\begin{figure}
    \centering
    \includegraphics[width=9cm]{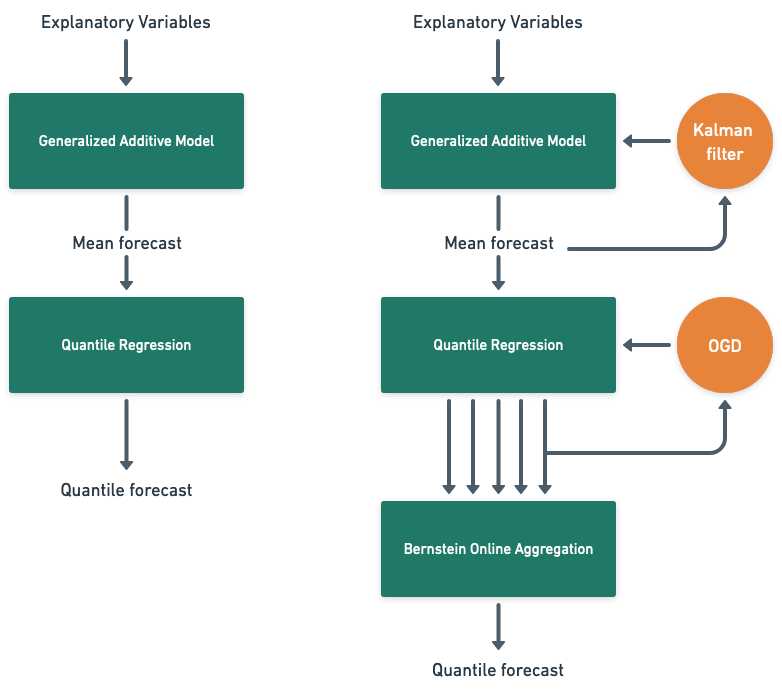}
    \caption{Our two-step procedure starts from explanatory variables, computes an intermediate mean forecast, and then estimate the quantile forecast. On the left we describe the offline model (Section \ref{sec:offline} and \cite{browell2021probabilistic}). Our contributions are represented by the right diagram, introducing our adaptive model (Sections \ref{sec:adaptgam} to \ref{sec:aggregation}), in which both steps are adapted.}
    \label{fig:schema}
\end{figure}

The methodology employed and our contributions are summarized in Figure~\ref{fig:schema}. We adopt a two-step procedure.
First, we employ a widely used model class, namely that of Generalized Additive Models (GAMs), to forecast the expected value of the variable of interest. This is a non-adaptive model fitted to several years of training data. We consider a state-space representation to obtain an adaptive variant, relying on the Kalman filter for the inference similarly to \cite{obst2021adaptive,de2022state}.
Second, we focus on forecasting the quantiles of the net-demand.
While the Kalman filter already models it as a Gaussian, we observe that this forecast is not always well-calibrated. Therefore, we use a set of quantile regressions on the residuals of the mean estimates. We apply the online gradient descent (OGD) to obtain adaptive quantile regressions. We select the best gradient step by exploiting an expert aggregation algorithm, Bernstein Online Aggregation \cite{wintenberger2017optimal}; it means we compute the OGD with several step sizes (multiple arrows in Figure~\ref{fig:schema}) and our final forecast is a combination of these forecasts; the quantile loss is optimized online as proposed by \cite{zaffran2022adaptive}.

In Section~\ref{sec:theory} we present adaptive methods for probabilistic forecasting, and we discuss their evaluation in Section~\ref{sec:evaluation}. Then we apply our approach on two data sets. In Section~\ref{sec:GB}, we consider the regional net-load in Great Britain, and we extend the work of \cite{browell2021probabilistic} to the adaptive setting, also extending the data set with more recent data (including the coronavirus crisis). Furthermore, we show in Section~\ref{sec:rmcap} that adaptive models have fewer needs for good explanatory variables; indeed, removing two (difficult to obtain) variables from the model has a lower impact on the adaptive variant. In Section~\ref{sec:US}, we apply the methods to the load of seven major cities in the United States \cite{ruan2020cross}. In the latter application, we predict daily, rather than half-hourly, data to determine whether our approach works well at this time granularity.

\section{Theoretical Framework}\label{sec:theory}
We aim to forecast a variable of interest $y_t\in\R$ given some explanatory variables $x_t\in\R^d,\ d>0$. We formally introduce the impact of time dependence on our approach. Let $\mathcal{F}_t=\sigma(x_1,y_1,\hdots, x_t,y_t)$ be the natural filtration; it models the information contained in the observations up to time $t$. We discriminate between two settings represented with two diagrams in Figure~\ref{fig:schema}:
\begin{itemize}
    \item
    {\it Offline} or {\it Batch} (Section~\ref{sec:offline}).
    The model is learned on a training period, for instance on data up to time $n_{\rm train}$. We estimate $\mathcal{L}(y_t\mid x_t, \mathcal{F}_{n_{\rm train}})$, the conditional distribution of $y_t$ given $x_t, \mathcal{F}_{n_{\rm train}}$.
    \item
    {\it Online} or {\it Adaptive} (Sections~\ref{sec:adaptgam} and~\ref{sec:adaptquantiles}).
    The model is learned sequentially. We estimate $\mathcal{L}(y_t\mid x_t, \mathcal{F}_{t-1})$ at each time $t$.
\end{itemize}

\subsection{Offline Model}\label{sec:offline}
Motivated by \cite{browell2021probabilistic}, we decompose our model into two steps. We forecast the conditional mean with a generalized additive model (GAM), and then quantile regressions on the GAM residuals yield quantile forecasts.
\begin{enumerate}
\item
We model $y_t$ as a Gaussian random variable whose mean is a sum of the effects of the covariates:
\begin{align}
    \label{eq:gam}
	y_t = \sum\limits_{j=1}^d f_j(x_{t,j}) + \varepsilon_t \,, \quad \varepsilon_t\sim\mathcal{N}(0,\sigma^2)\,,
\end{align}
where the effects $f_1,\hdots f_d$ are either linear or nonlinear. In the latter case, the effects are built using linear combinations of spline bases \cite{wood2017generalized}. The optimization of these effects on the training set yields a mean forecast $\hat y_t = \mathbb{E}[y_t\mid x_t, \mathcal{F}_{n_{\rm train}}]$.

\item
The Gaussian assumption with fixed variance is violated in practice. Therefore, we fit a set of quantile regressions \cite{koenker1978regression} on the residuals to predict the distribution around the mean. We use an intermediate vector of covariates $z_t\in\R^{d_0},\ d_0>0$ derived from the original vector $x_t$; depending on the application, $z_t$ may contain the GAM prediction, the GAM effects $f_j(x_{t,j})$ ... For any quantile level $q$, we define a vector $\beta_q\in\R^{d_0}$ via the following minimization problem:
\begin{align}
    \label{eq:quantreg}
	& \beta_q \in \arg\min\limits_{\beta\in\R^{d_0}} \sum\limits_{t=1}^{n_{\rm train}} \rho_q(y_t-\hat y_t, \beta^\top z_t) \,, \\
	\label{eq:pinball}
	& \rho_q(y,\hat y_q)=(\mathds{1}_{y<\hat y_q} - q)\ (\hat y_q-y) \,.
\end{align}
Then, we predict the quantile for probability level $q$ with $\hat y_t + \beta_q^\top z_t$.
This is justified by the following well-known lemma~\cite{koenker2001quantile}.
\end{enumerate}
\begin{lemma}
	Let $Y$ be an integrable real-valued random variable. For any $0<q<1$ the $q$-quantile of $Y$ denoted by $Y_q$ satisfies $Y_q\in\arg\min\mathbb{E}[\rho_q(Y,Y_q)]$.
\end{lemma}
This two-step procedure is motivated by computational time and by the fact that the data is sparse in the tails (sparser than around the mean) as in \cite{gaillard2016additive}. Indeed, we could also use quantile GAM~\cite{fasiolo2021fast}, but it is more time-consuming than simple quantile regressions. Hence the need to use simple linear adjustments to correct the (more richly-parametrized) mean GAM. 
Also, we explain in the following paragraph that the mean GAM may be adapted using the Kalman filter, and that is not the case for quantile models, for which we adopt a different strategy, see~\ref{sec:adaptquantiles}.

The first natural adaptive forecaster consists in re-training the offline model at each time step. We call this procedure {\it incremental offline}.

\subsection{Adaptation of Generalized Additive Model}\label{sec:adaptgam}
We adapt the GAM as done by \cite{obst2021adaptive}. 
More precisely, we freeze the nonlinear effects learned on the training set (Section~\ref{sec:offline}); with these effects, we define a new covariate vector $f(x_t)=(\overline f_1(x_{t,1}),\hdots \overline f_d(x_{t,d}),1)^\top$, where $\overline f_j(x_{t,j})$ is the standardized version of $f_j(x_{t,j})$. Then we consider the linear Gaussian state-space model:
\begin{align}
    & \theta_t - \theta_{t-1} \sim\mathcal{N}(0,Q) \,, \\
    \label{eq:space} & y_t - \theta_t^\top f(x_t) \sim \mathcal{N}(0,\sigma^2) \,,
\end{align}
where $Q$ is the state noise covariance matrix, and $\sigma^2$ the space noise variance. In the preceding set of equations, we assume implicitly that the noises are independent.
Estimation in a linear Gaussian state-space model with known variances has been optimally solved by \cite{kalman1961new}:
\begin{theorem}[Kalman Filter]\label{th:kf}
Provided that the data-generating process is the state-space model with variances $Q$ and $\sigma^2$, and if the prior distribution of $\theta_1$ is $\mathcal{N}(\hat\theta_1,P_1)$, then the posterior distribution of the state is a Gaussian whose mean and covariance matrix have analytical forms. Precisely, we have $\theta_t\mid \mathcal{F}_{t-1}\sim\mathcal{N}(\hat\theta_t,P_t)$ with:
\begin{align}
    & P_{t\mid t} = P_t - \frac{P_t f(x_t)f(x_t)^\top P_t}{f(x_t)^\top P_t f(x_t)+\sigma^2} \,, \\
    \label{eq:updatetheta}
    & \hat{\theta}_{t+1} = \hat{\theta}_t - \frac{P_{t\mid t}}{\sigma^2} \Big(f(x_t) (\hat{\theta}_t^\top f(x_t)-y_t)\Big) \,, \\
    & P_{t+1} = P_{t\mid t} + Q \,.
\end{align}
\end{theorem}

The unsolved issue of linear Gaussian state-space models concerns the estimation of its variances $Q$ and $\sigma^2$.
In our setting, as well as in many applications, the variances are unknown. The choice of the variances can be seen as a parametrization of a gradient algorithm. Indeed, $f(x_t) (\hat{\theta}_t^\top f(x_t)-y_t)$ is the gradient of the quadratic loss with respect to $\theta$; therefore, the update \eqref{eq:updatetheta} is a gradient step with a {\it pre-conditioning} matrix $P_{t\mid t}/\sigma^2$ which depends crucially on the choice of $\sigma^2$ and $Q$. We propose different settings:
\begin{itemize}
    \item
    {\it Static}. We set $Q=0$ (constant state vector), and $\sigma^2=1$. This yields a detailed link with the gradient community \cite{de2021stochastic}; the constant state is equivalent to the i.i.d. assumption of gradient analyses in the setting $P_{t\mid t}\rightarrow 0$ (annealing step size).
    \item
    {\it Dynamic}. The natural approach aims at maximizing the likelihood; in that vein, we apply an iterative greedy procedure implemented in the R package \texttt{viking} \cite{viking}, that was applied on electricity load forecasting by \cite{obst2021adaptive,de2022state}. It yields a sparse matrix $Q$ assumed diagonal. In that setting, the gradient step does not converge to 0.
    Note that we apply this procedure to maximize the likelihood on a training set; therefore the variances $\sigma^2$ and $Q$ are learned in a {\it batch} fashion, but the underlying model assumes a time-varying state $\theta_t$ leading to an {\it adaptive} $\hat\theta_t$.
\end{itemize}

\subsection{Adaptation of the Quantile Forecasts}\label{sec:adaptquantiles}
The novelty of the present approach lies in the adaptation of probabilistic forecasts. We compare different methods.

\subsubsection{Gaussian Posterior from Kalman}
We first recall that the Kalman filter does not only yield a mean forecast but already a probabilistic forecast. Indeed, at each time step $t$, Theorem \ref{th:kf} yields $\mathcal{L}(\theta_t\mid \mathcal{F}_{t-1})=\mathcal{N}(\hat\theta_t, P_t)$. 
From the state posterior distribution and the observation distribution~\eqref{eq:space}, we deduce 
\begin{align}
    \mathcal{L}(y_t \mid x_t, \mathcal{F}_{t-1}) = \mathcal{N}(\hat\theta_t^\top f(x_t), f(x_t)^\top P_tf(x_t)+\sigma^2)\,.
\end{align}
It yields a readily computable probabilistic forecast; for any quantile level $q$, denoting by $N_q$ the quantile of the standard normal distribution, we predict the $q$-quantile of $y_t$ as
\begin{align}
    \hat y_{t,q} = \hat\theta_t^\top f(x_t) + N_q \sqrt{f(x_t)^\top P_tf(x_t)+\sigma^2} \,.
\end{align}
However, this first adaptive forecaster would be an adaptive variant of the offline Gaussian distribution on the offline GAM residuals, where only the GAM is adapted but the variance of the Gaussian distribution is still fixed. The quantile regressions were specifically introduced because the Gaussian assumption with fixed variance is violated in practice. We see in the experiments that this Gaussian forecaster is not always well-calibrated.

\subsubsection{Quantile Regressions on Kalman Residuals}
We then remark that the offline quantile regressions should work well in the case where the conditional distribution of the residual $y_t-\hat y_t$ given the covariates $z_t$ is fixed. The need for adaptive quantile forecasts is motivated only by changes in the residual distribution.
Yet a critical property of the Kalman filter is that the residuals are stationary, provided that the state-space model is well-specified.
Consequently, the dependence of the Kalman residuals on the quantile covariates should be more stable than that of the offline GAM.
Therefore, we combine state-space adaptation of the GAM and offline quantile regressions.

\subsubsection{Online Gradient Descent on the Pinball Loss}
We consider an online quantile regression applying online gradient descent (OGD) on the pinball loss.
Precisely, we defined in Section~\ref{sec:offline} the offline quantile regression for the $q$-quantile with a vector of covariates $z_t\in\R^{d_0},\ d_0>0$:
\begin{align}
    \beta_q \in \arg\min_{\beta\in\R^{d_0}} \sum\limits_{t=1}^{n_{\rm train}} \rho_q(y_t-\hat y_t, \beta^\top z_t)\,.
\end{align}
Motivated by the gradient interpretation of the Kalman filter (Section~\ref{sec:adaptgam}), we apply the OGD to estimate recursively a vector $\beta_{t,q}$. We start from any $\beta_{1,q}\in\R^{d_0}$ and at each step we update it with a step in the direction opposite to the gradient of the loss:
\begin{align}
	\beta_{t+1,q} = \beta_{t,q} - \alpha \frac{\partial \rho_q(y_t-\hat y_t, \beta^\top z_t)}{\partial \beta}\Big|_{\beta_{t,q}} \,.
\end{align}
The gradient step size $\alpha$ is the important parameter, analogous to the variances that are determining the gradient step in the gradient interpretation of the Kalman filter.
We use a constant $\alpha$ in the OGD, and we standardize the covariates $z_t$. Note that the gradient is not well defined for the singular point $y_t=\hat y_t+ \beta_{t,q}^\top z_t$; we choose to set it to $0$ in the remote case that this happens in practice.

\subsubsection{Bernstein Online Aggregation to Choose the Step Size}\label{sec:aggregation}
In order to choose the step size in the previous method, we rely on aggregation of experts as proposed by \cite{zaffran2022adaptive} and similarly by \cite{vanerven2021}. Our procedure is run separately for each quantile level $q$ using the R package \texttt{opera} \cite{opera}.
First, we run the OGD with different possible values $(\alpha_k)_{1\le k\le K}$; at each time $t$, it yields a forecast $\hat y_{t,q}^{(k)}$ for each step size $\alpha_k$.
Second, we combine these forecasts using Bernstein Online Aggregation \cite{wintenberger2017optimal}; the principle of aggregation is to forecast $\hat y_{t,q}=\sum_{k=1}^K p_{t,q}^{(k)} \hat y_{t,q}^{(k)}$, where the weights $p_{t,q}^{(k)}$ are obtained sequentially.
The properties obtained by the online learning literature guarantee that the total pinball loss of the aggregation has a small regret compared with the total pinball loss of the best expert. A similar approach was applied by \cite{berrisch2021crps}, where the aggregation weights are estimated jointly for all quantiles; the authors assume the weights are smoothed functions of the quantile level and they optimize directly the CRPS instead of point quantile functions.

\subsection{Computational Complexity of the Methodology}
The computations of the Kalman filter are separated in two. The estimation of the variances in the dynamic setting is costly (more costly than a single GAM estimation). However, the updates detailed in Theorem~\ref{th:kf} are very efficient and their computation on the whole data set is lighter than a single GAM estimation.
On the other hand, the cost of the {\it incremental offline} GAM is much bigger because each day, we have a GAM estimation. Therefore, the cost of the mean adaptive mean forecast is lower than that of the {\it incremental offline} GAM, and is concentrated at the calibration of the variances.

Considering our probabilistic forecasts, note that the OGD and the BOA algorithm are also obtained with a single pass on the data, therefore they are much lighter than re-estimating quantile regressions every day.

Numerical values of computational times are given in the supplementary material \cite{joseph_de_vilmarest_2022_7849665}.

\section{Evaluation}\label{sec:evaluation}
For both mean and probabilistic forecasting tasks, we evaluate the forecasts qualitatively as well as quantitatively.
\subsection{Mean Forecast}
We evaluate through the root-mean-square-error (RMSE) and the mean absolute error (MAE), defined on a test set $\mathcal{T}$ by
\begin{align}
    & \text{RMSE} = \sqrt{\frac{1}{|\mathcal{T}|}\sum\limits_{t\in\mathcal{T}} (y_t-\hat y_t)^2} \,,\\
    & \text{MAE} = \frac{1}{|\mathcal{T}|}\sum\limits_{t\in\mathcal{T}} |y_t-\hat y_t| \,.
\end{align}

Forecasting the mean of the variable of interest is a way to minimize the quadratic loss, thus the RMSE is natural. The MAE is known to be more robust to outliers.

We don't use the mean absolute percentage error (MAPE) because in the case of the net-load the variable may be close to 0 or even negative, see Section~\ref{sec:GBoffline}.
However, our data sets are composed of different time series to predict and to obtain global evaluation we use relative metrics. We choose to divide by the same metric applied to the mean.

Formally, we have $N$ time series $(y_{t,i})_{t\in\mathcal{T}, 1\le i\le N}$ and corresponding forecasts $(\hat y_{t,i})_{t\in\mathcal{T}, 1\le i\le N}$. We estimate the means $\overline y_i=\frac{1}{|\mathcal{T}|}\sum_{t\in\mathcal{T}} y_{t,i}$. Our aggregate metrics are defined by
\begin{align}
    & \text{nRMSE} = \sqrt{\frac1N\sum\limits_{1\le i\le N} \frac{\sum\limits_{t\in\mathcal{T}} (y_{t,i}-\hat y_{t,i})^2}{\sum\limits_{t\in\mathcal{T}} (y_{t,i}-\overline y_i)^2}} \,,\\ 
    & \text{nMAE} = \frac1N\sum\limits_{1\le i\le N} \frac{\sum\limits_{t\in\mathcal{T}} |y_{t,i}-\hat y_{t,i}|}{\sum\limits_{t\in\mathcal{T}} |y_{t,i}-\overline y_i|} \,.
\end{align}
Our normalized metrics may be interpreted as unexplained variations, in the opposite way of the R-squared.

\subsection{Probabilistic Forecast}
There are different ways to evaluate quantile forecasts. However, a necessary condition for the forecast to be meaningful is {\it reliability}, also known as {\it calibration}. A quantile forecast is reliable if the observed frequency of exceedance coincides with the quantile level. The forecast of a $q$-quantile is expected to be empirically higher than the quantity of interest for a fraction $q$ of the data set and smaller for a fraction $1-q$. Therefore, we compare the observed frequencies with the quantiles in reliability diagrams. 

Reliable forecasts may then be ranked by sharpness.
Numerical evaluation is obtained by the pinball loss $\rho_q$ defined by Equation \eqref{eq:pinball}.
A way to combine the pinball losses at different quantile levels is to use the continuous ranked probability score (CRPS) \cite{gneiting2007strictly}, defined equivalently by the two following expressions:
\begin{align}
    \nonumber
	\text{CRPS}(F,y) & = \int\limits_{-\infty}^{+\infty} (F(x)-\mathds{1}_{y\le x})^2 dx \\
	\label{eq:crps}
	& = 2 \int\limits_0^1 \rho_q(y, F^{-1}(q)) dq \,,
\end{align}
where $y$ is the observation and $F$ the predicted cumulative distribution function. Remark that for any finite $y$ we have $\rho_0(y, F^{-1}(0))=\rho_1(y, F^{-1}(1))=0$, therefore the CRPS is not a good performance indicator for the tail estimation.

We use the discrete approximation of the integral in Equation~\eqref{eq:crps}. We have a set of forecasted quantiles $\hat y_{q_1}, \hdots , \hat y_{q_l}$, and we define the RPS as the integral of the piecewise linear function interpolating $0,\rho_{q_1}(y,\hat y_{q_1}), \hdots \rho_{q_l}(y,\hat y_{q_l}),0$ at the points $0,q_1,\hdots q_l,1$. This yields
\begin{align}
	\text{RPS}\big((\hat y_{q_1}, \hdots , \hat y_{q_l}),y\big) = \sum\limits_{i=1}^l \rho_{q_i}(y,\hat y_{q_i}) (q_{i+1}-q_{i-1}),
\end{align}
where we define $q_0=0,q_{l+1}=1$. Then we simply use the RPS averaged over time on a test set.

As for mean evaluation, we then aggregate this metric over the multiple time series (regions in \ref{sec:GB}, cities in \ref{sec:US}). We divide by the mean absolute error of the mean, which is also the CRPS of a Dirac distribution centered at the mean:
\begin{align}
    \text{nRPS} = \frac1N\sum\limits_{1\le i\le N} \frac{\sum\limits_{t\in\mathcal{T}} RPS\big((\hat y_{t,i,q_1}, \hdots , \hat y_{t,i,q_l}),y_{t,i}\big)}{\sum\limits_{t\in\mathcal{T}} |y_{t,i}-\overline y_i|}. 
\end{align}

\section{Regional Net-Load in Great Britain}\label{sec:GB}
We first study the data set created by \cite{browell2021probabilistic}. They studied the data from 2014 to 2018, and we augment the period to range from 2014 to 2021 \cite{joseph_de_vilmarest_2022_7849665}, allowing us to integrate the unstable covid period.
We refer to \cite{browell2021probabilistic} and \cite{jethro_browell_2021_5031704} for more details on the data set. Due to Brexit, the day-ahead electricity price definition changed; we decided to remove this explanatory variable from the data set.

\subsection{Data Presentation and Offline Model}\label{sec:GBoffline}
\begin{figure*}
\centering
\includegraphics[width=6.5cm]{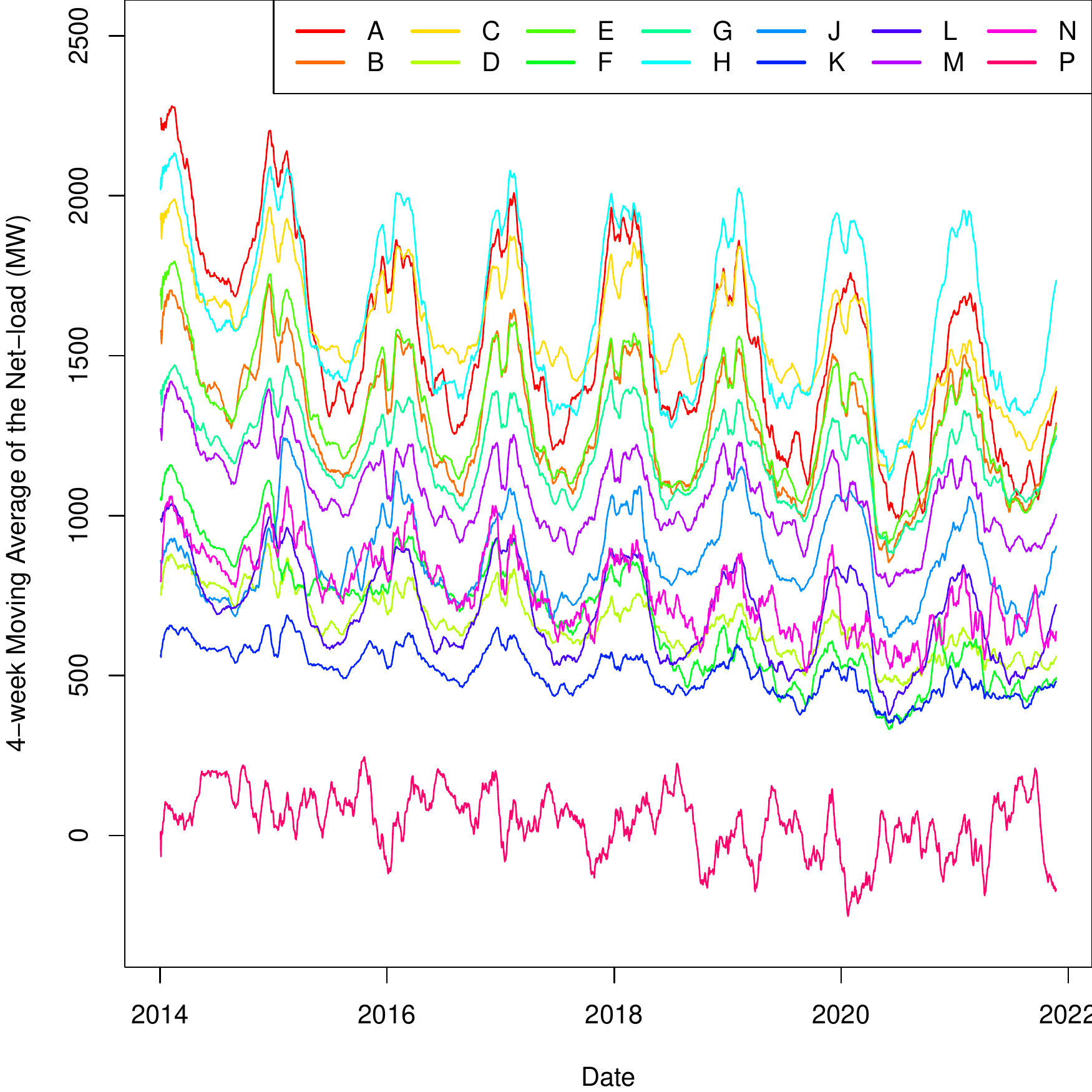}
\hspace{0.5cm}
\includegraphics[width=6.5cm]{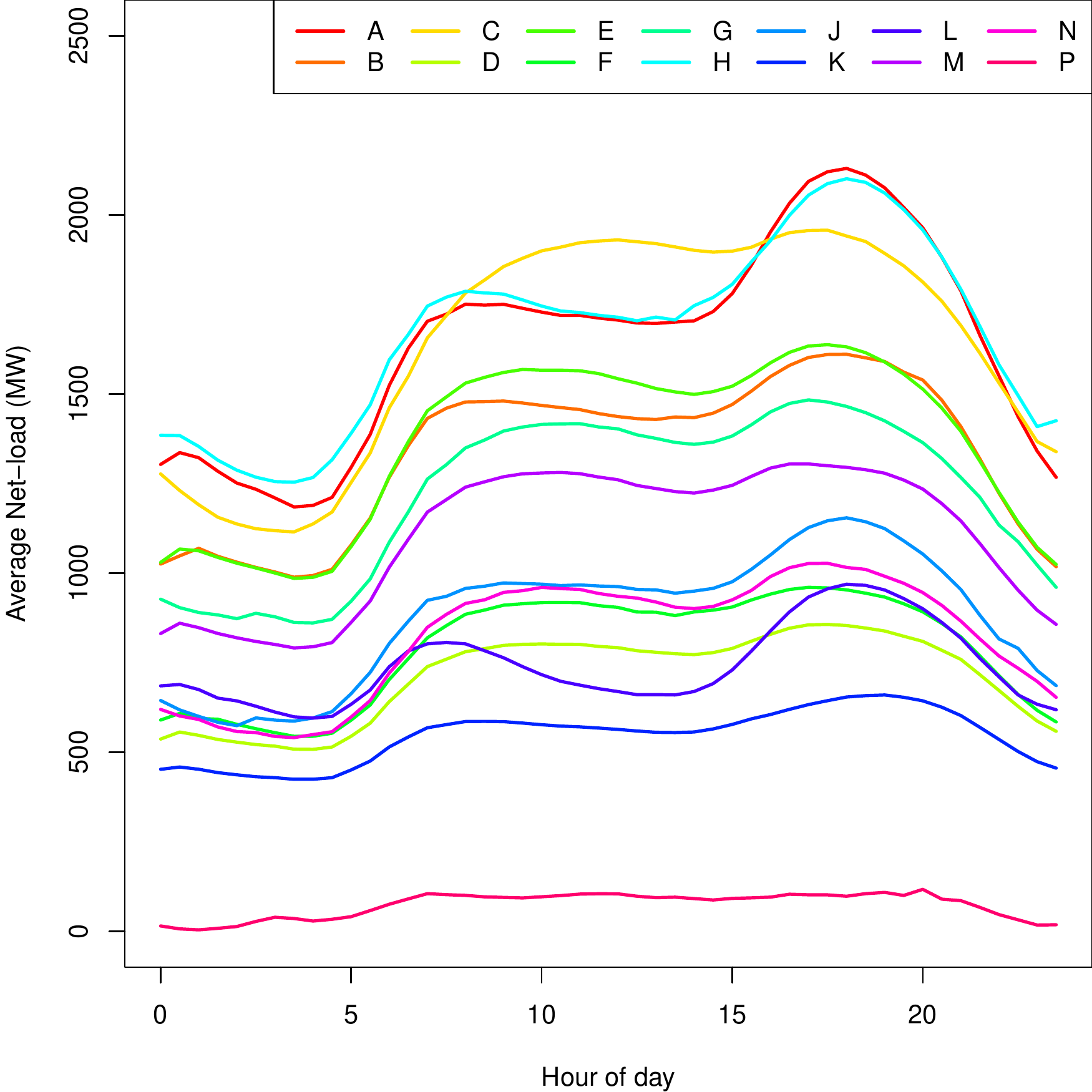}
\caption{On the left: evolution of the net-load of the 14 regions. On the right: daily profiles. We observe that in region P (North Scotland) the embedded generation often exceeds the consumption and the daily profile is close to 0. Also, this high level of renewables implies higher weather-driven volatility, and the net-load does not have a clear yearly profile as in the other regions.}
\label{fig:GBload}
\end{figure*}

We are interested in forecasting the electricity net-load, defined as the difference between electricity consumption and embedded generation (mostly wind and solar production). We consider the regional data from Great Britain. Great Britain is divided into 14 regions called Grid Supply Point Groups. The time granularity is half an hour and we assume a one-day delay for data availability: each day at midnight we can update the model with data up to 24 hours before. We display in Figure~\ref{fig:GBload} the evolution of the net-load in each region, as well as the daily profiles. We observe different behaviors. In particular, region P (North Scotland) often has a negative net-load, meaning the consumption is smaller than the embedded production.

For each region independently, the model designed by \cite{browell2021probabilistic} to predict quantiles of the net-load is decomposed into three steps. First, a GAM is fitted to forecast the mean. Second, a set of quantile regressions are fitted to the GAM residuals (between $2.5\%$ and $97.5\%$). Third, extreme quantiles are modeled via a generalized Pareto distribution. The first two match our framework introduced in Section~\ref{sec:offline}; we apply the same additive model for the mean, then the same quantile regression.

\begin{enumerate}
\item
We use the formula of the {\it GAM Point} of \cite{browell2021probabilistic}, where electricity prices are removed.
The normalized electricity net-load is modeled by \eqref{eq:gam} with the following covariates: a trend, the time of year, the time of day, the day of the week, school holidays, a moving average of the net-load with one-day delay, the temperature at maximum population density, a moving average of that temperature, solar radiation combined with the embedded solar generation capacity, the wind speed combined with the embedded wind generation capacity, and the precipitation.

Weather variables are weather forecasts. All nonlinear effects are built using cubic regression splines.

\item
For the quantile regressions, we also use the model of \cite{browell2021probabilistic}: the GAM residuals are modeled via linear functions of the GAM prediction, the squared prediction, the product of solar radiation and embedded solar capacity, the wind speed, the temperature, as well as of the categorical versions of the time of day and day of the week. For these two latter variables, we have an additive constant defined for each value of the categorical variable.
\end{enumerate}

Note that the GAM and the quantile regressions are not trained per half an hour; they directly incorporate the time of day as a covariate.

\subsection{Performances of Mean Forecast}
\begin{figure*}
    \centering
    \includegraphics[width=5.4cm]{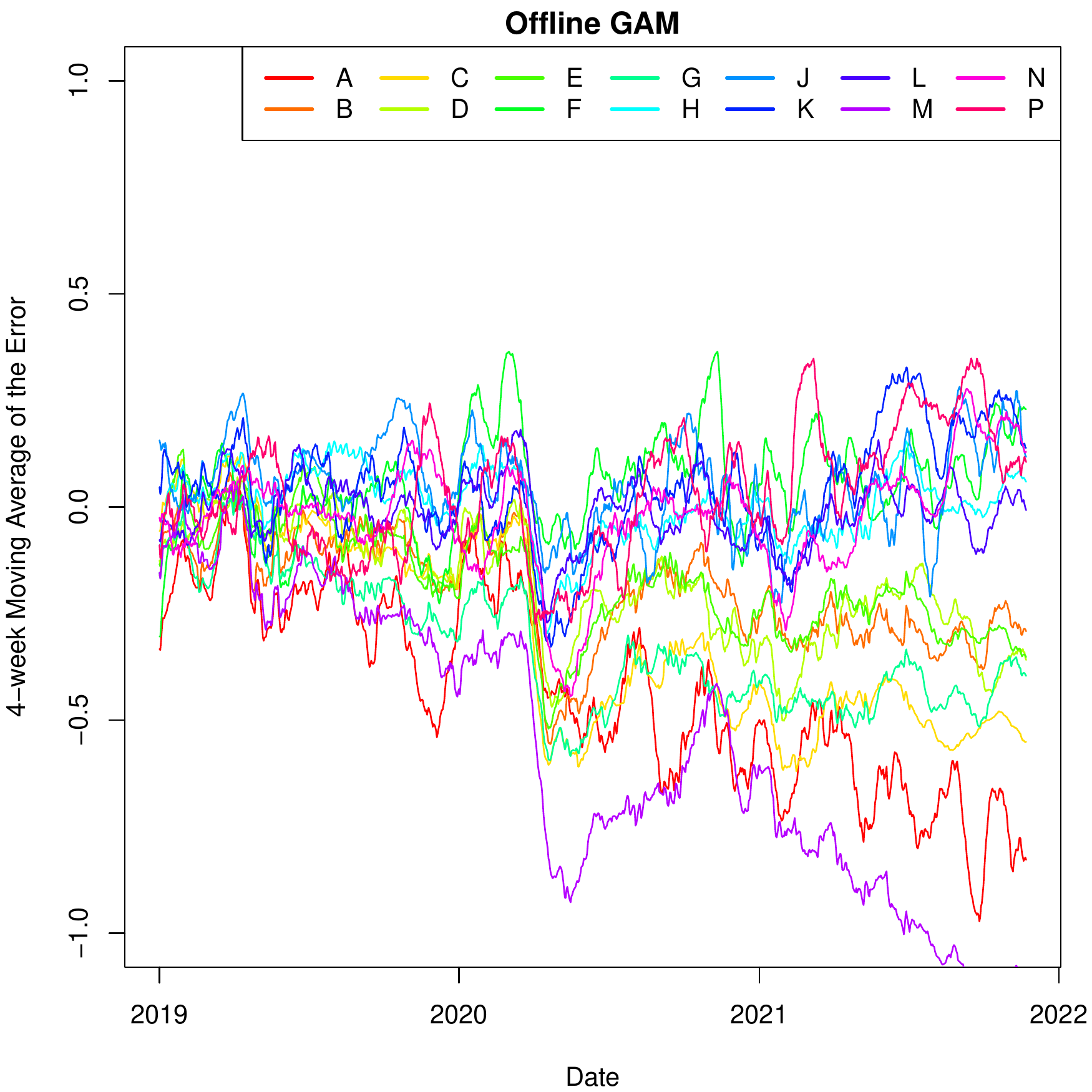}
\hspace{0.5cm}
    \includegraphics[width=5.4cm]{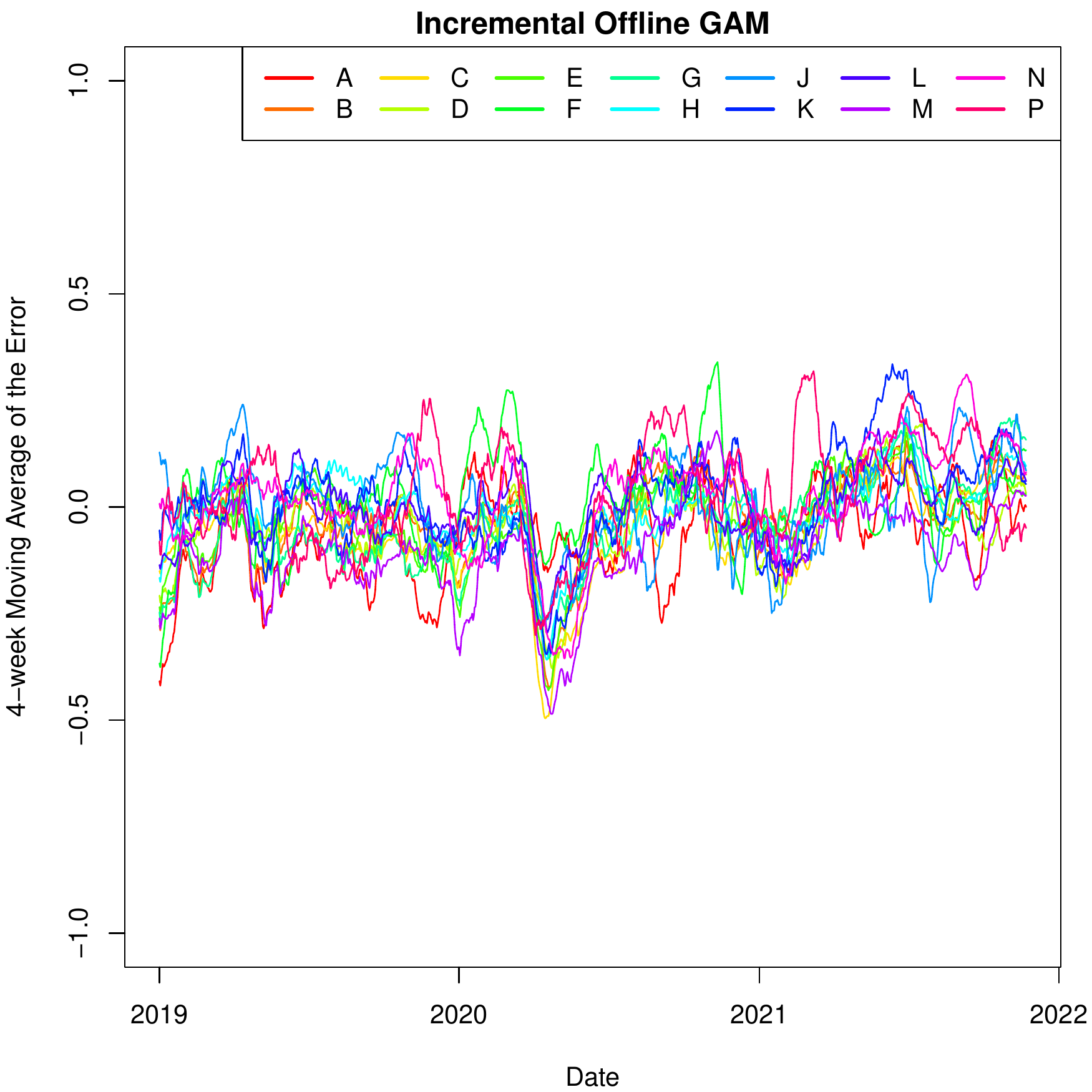}
\hspace{0.5cm}
    \includegraphics[width=5.4cm]{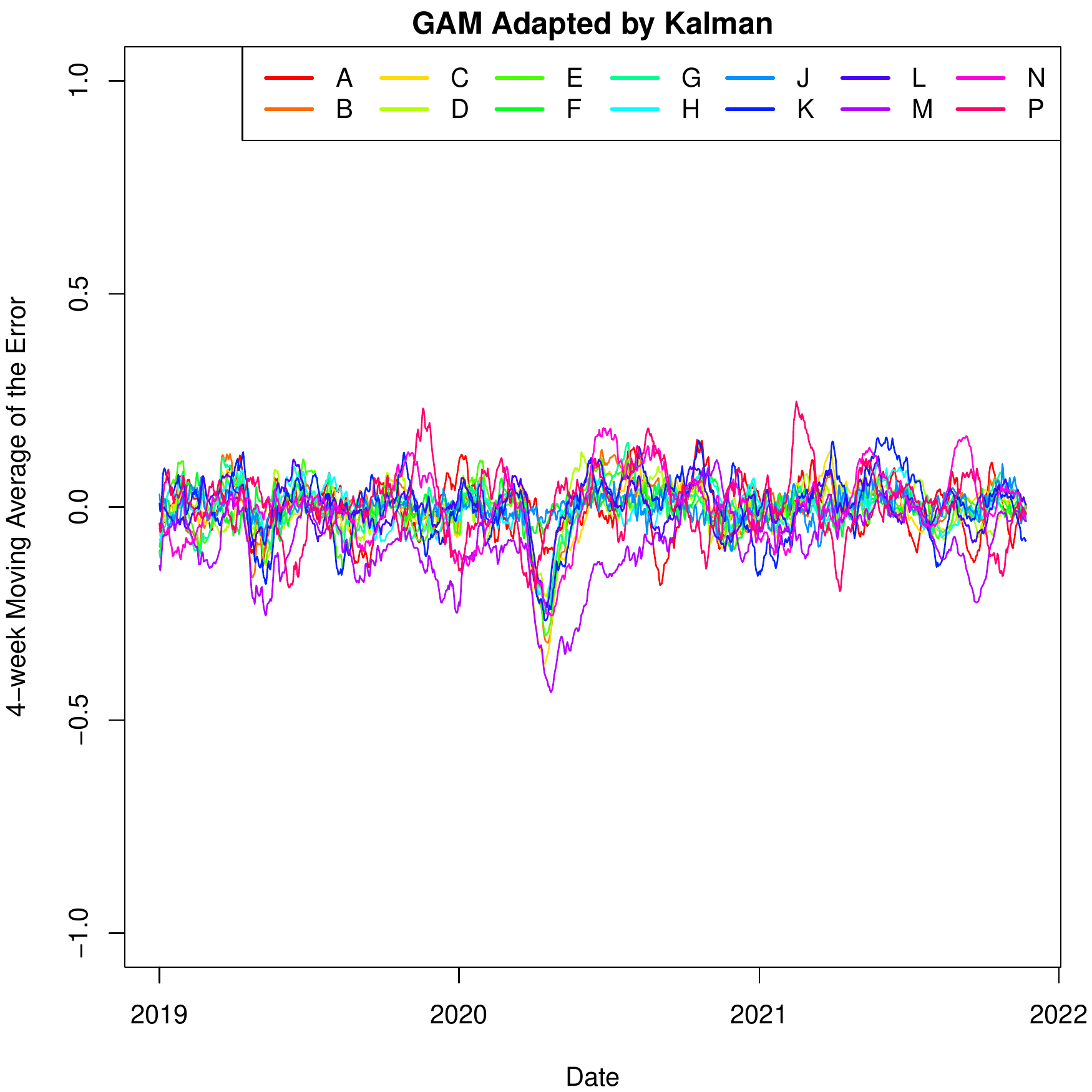}
    \caption{Evolution of the standardized error for the offline model trained on data up to December 31, 2018 (left), the incremental offline model trained each day (middle), and the GAM adapted by the Kalman filter (right).}
    \label{fig:GBerrorevol}
\end{figure*}

\begin{figure*}
\centering{
\includegraphics[width=5.4cm]{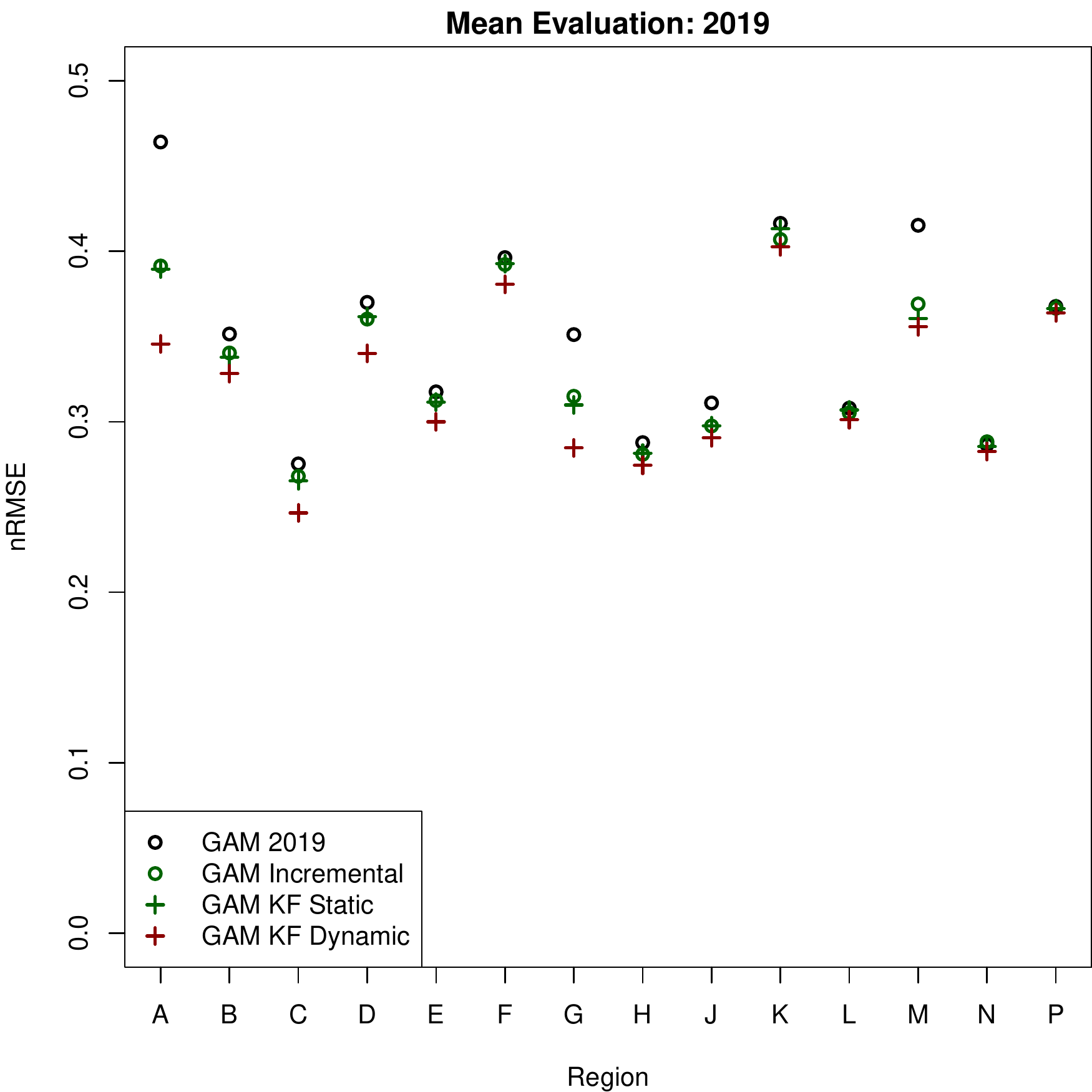}
\hspace{0.5cm}
\includegraphics[width=5.4cm]{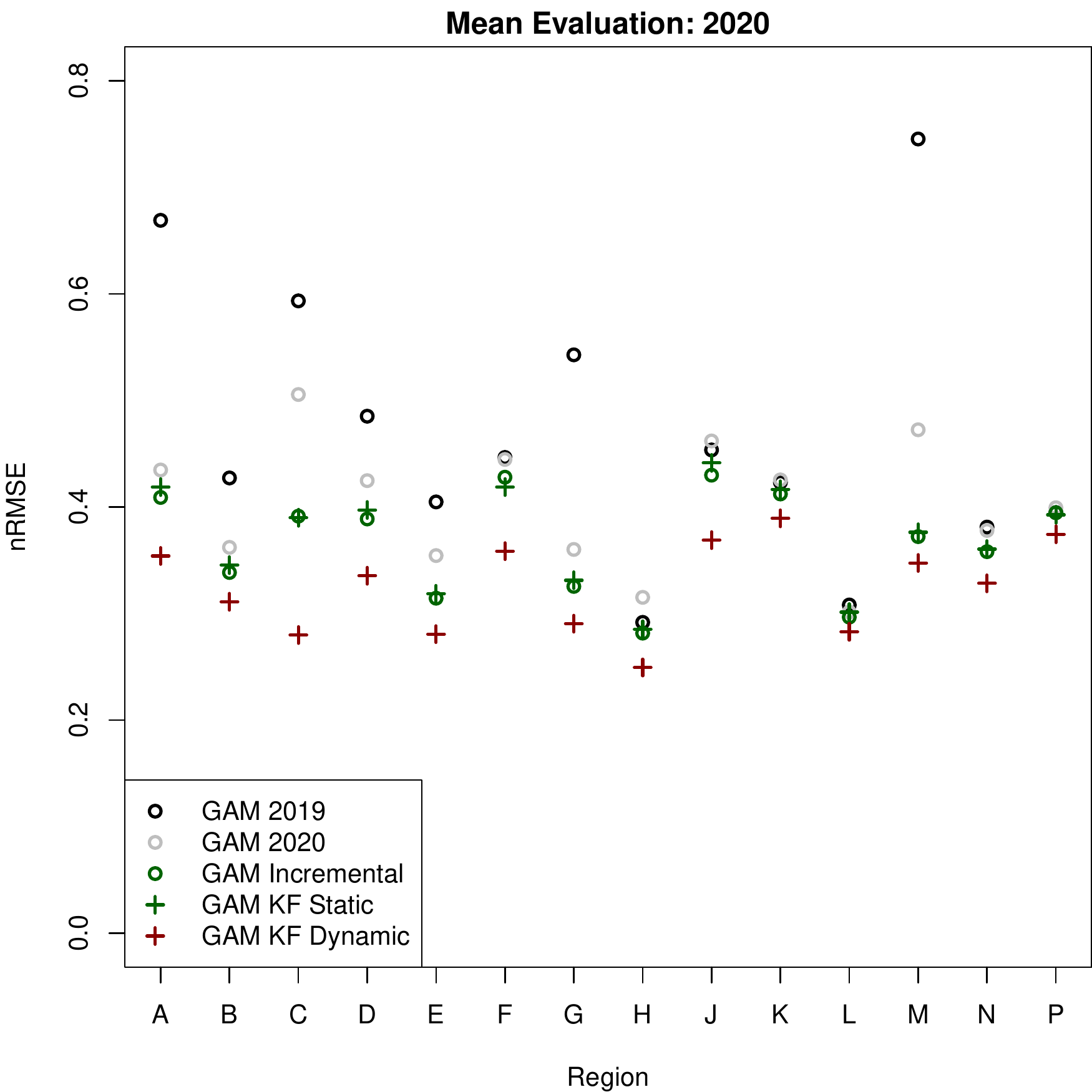}
\hspace{0.5cm}
\includegraphics[width=5.4cm]{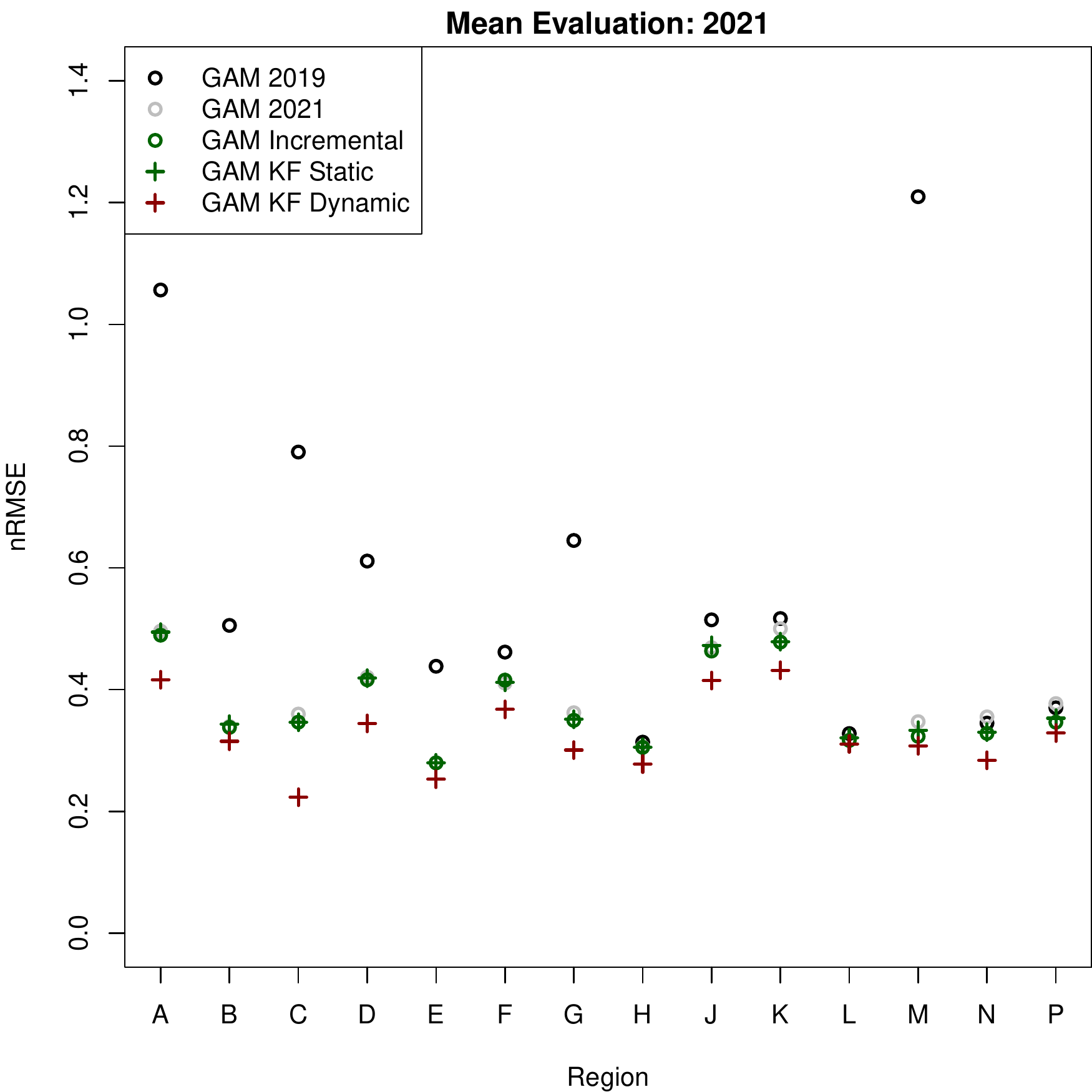}}
\caption{nRMSE for each region in Great Britain. We compare the offline GAM to the incremental offline GAM (trained each day on incremental training data), the Kalman filter in the static setting (degenerate covariance matrix of the state noise $Q=0$), and the Kalman filter in the dynamic setting. We divide the test set in three (2019, 2020 and 2021). For the last two years, we compare also with the GAM re-trained each year: GAM $y$ is the GAM trained with the data up to year $y-1$ included.}
\label{fig:rmse_gb}
\end{figure*}

We display the evolution of the error in Figure~\ref{fig:GBerrorevol}.
We improve the mean forecast performance in almost all regions and all years using state-space adaptation, considering the nRMSE, see Figure~\ref{fig:rmse_gb}.
We display the aggregate nRMSE and nMAE in Table~\ref{tab:GBrmse} for different methods and different years. When we compare the dynamic Kalman filter to the incremental offline GAM (model re-trained each day), we reduce the nRMSE by approximately $4\%$ in 2019, $7\%$ in 2020, and $8\%$ in 2021; furthermore, we have a much lower computational cost per day.
\begin{table*}
\caption{Aggregate metrics on the 14 regions for the different models. We compare to persistence benchmarks (lags of the variable of interest).}
\label{tab:GBrmse}
\centering
\begin{tabular}{c|c c|c c|c c}
& \multicolumn{2}{c|}{2019} & \multicolumn{2}{c|}{2020} & \multicolumn{2}{c}{2021} \\
\hline
Forecast & nRMSE & nMAE & nRMSE & nMAE & nRMSE & nMAE \\
\hline
Persistence (7 days) & 0.691 & 0.589 & 0.710 & 0.599 & 0.737 & 0.639 \\
Persistence (2 days) & 0.767 & 0.686 & 0.755 & 0.668 & 0.736 & 0.668 \\
\hline
Offline GAM & 0.356 & 0.327 & 0.485 & 0.453 & 0.635 & 0.601 \\
Incremental offline GAM (yearly) & - & - & 0.407 & 0.376 & 0.387 & 0.378 \\
Incremental offline GAM (daily) & 0.338 & 0.307  & 0.370 & 0.344 & 0.377 & 0.365 \\
\hline
Kalman GAM (Static) & 0.337 & 0.307& 0.374  & 0.347 & 0.380 & 0.368 \\
Kalman GAM (Dynamic) & {\bf 0.324} & {\bf 0.292} & {\bf 0.328} & {\bf 0.301} & {\bf 0.332} & {\bf 0.307}
\end{tabular}
\end{table*}

\subsection{Performances of Quantile Forecast}
\begin{figure*}
\centering
\includegraphics[width=5.4cm]{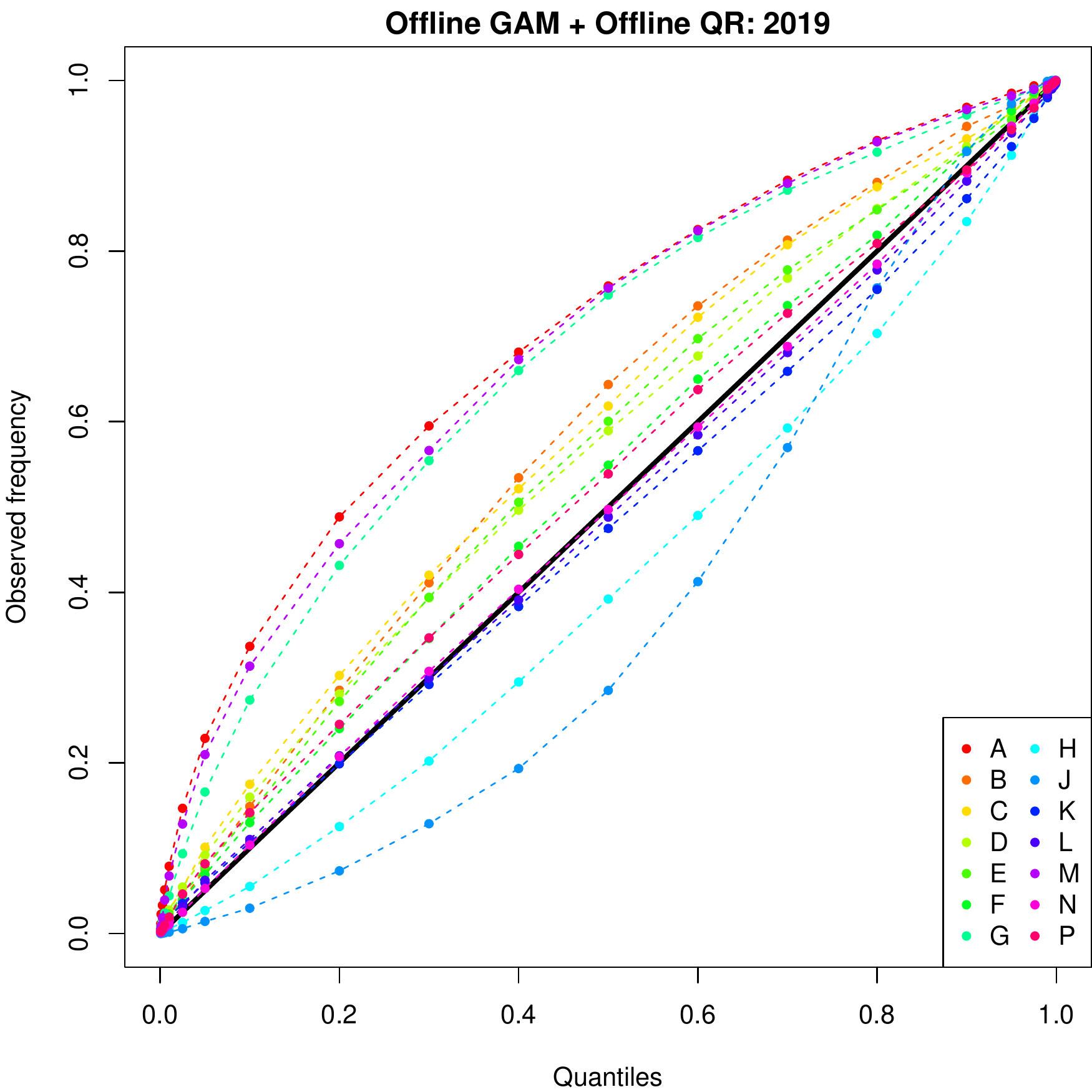}
\hspace{0.5cm}
\includegraphics[width=5.4cm]{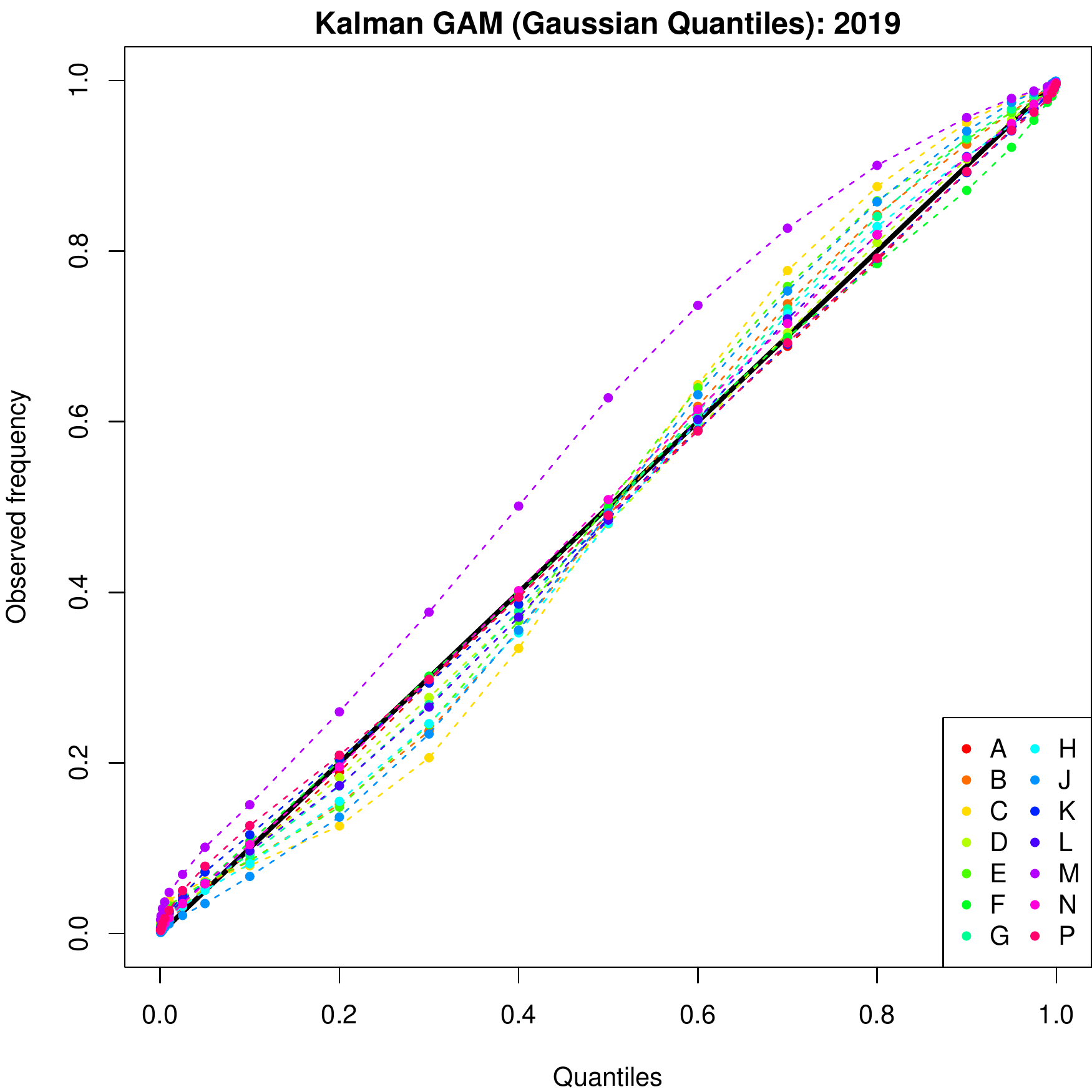}
\hspace{0.5cm}
\includegraphics[width=5.4cm]{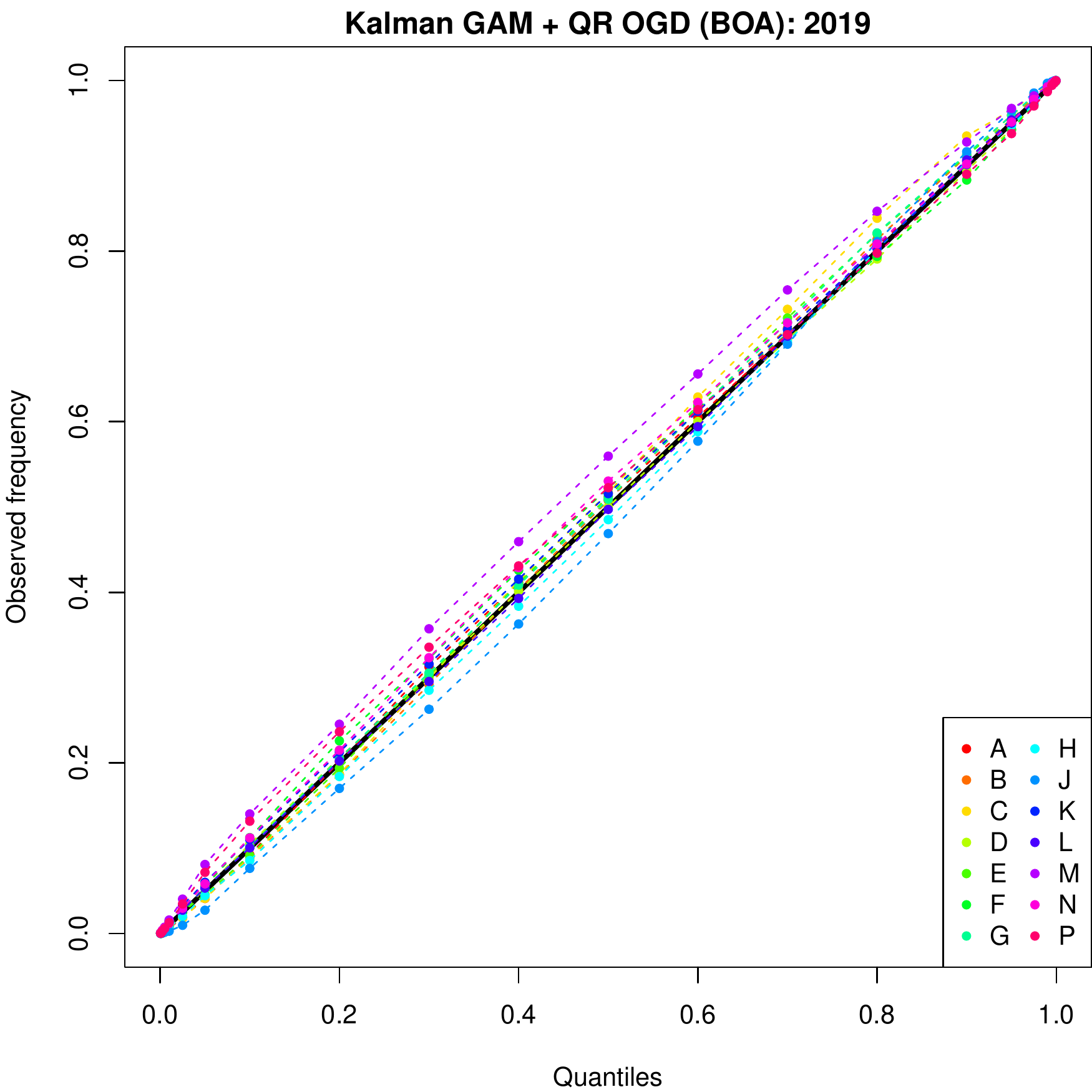}
\includegraphics[width=5.4cm]{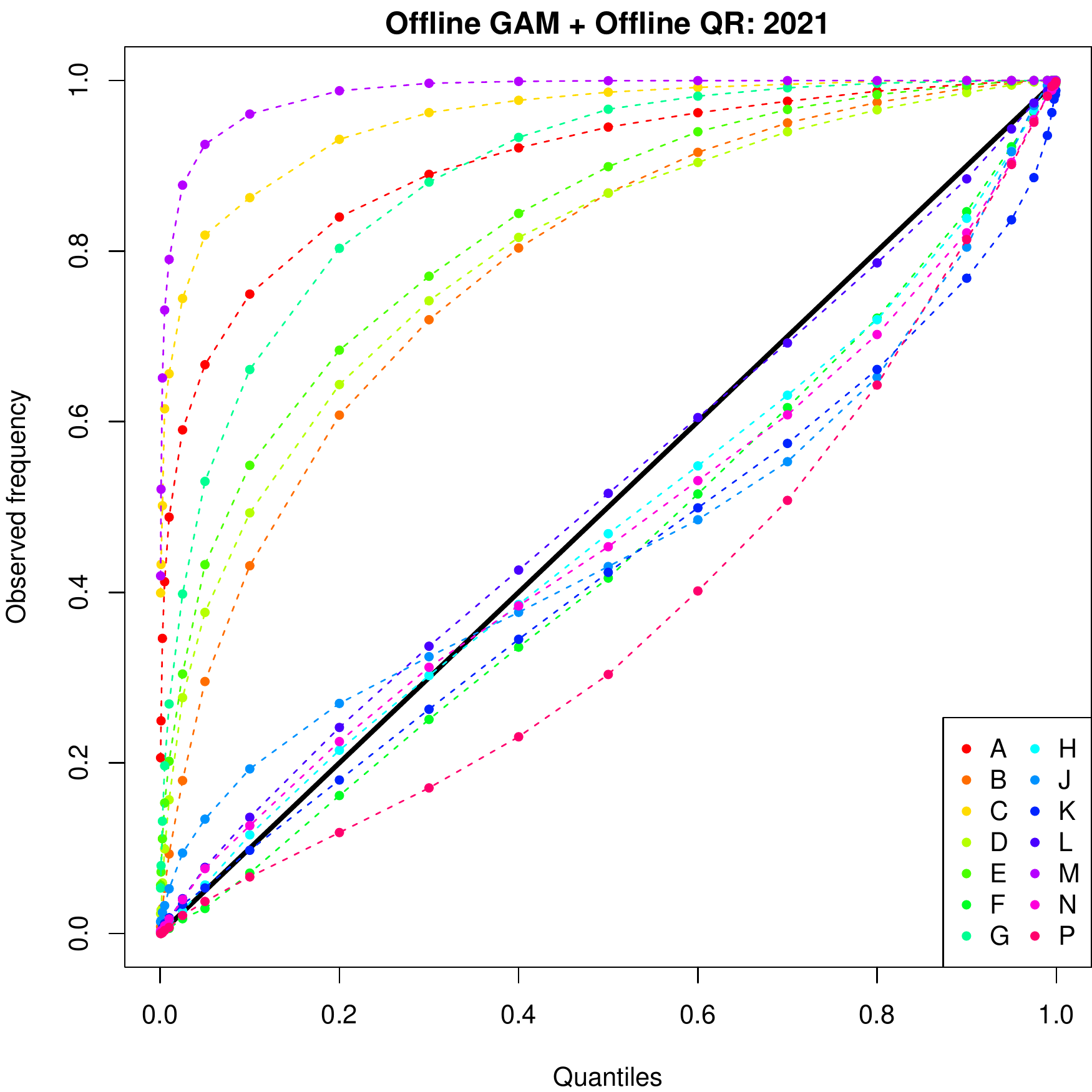}
\hspace{0.5cm}
\includegraphics[width=5.4cm]{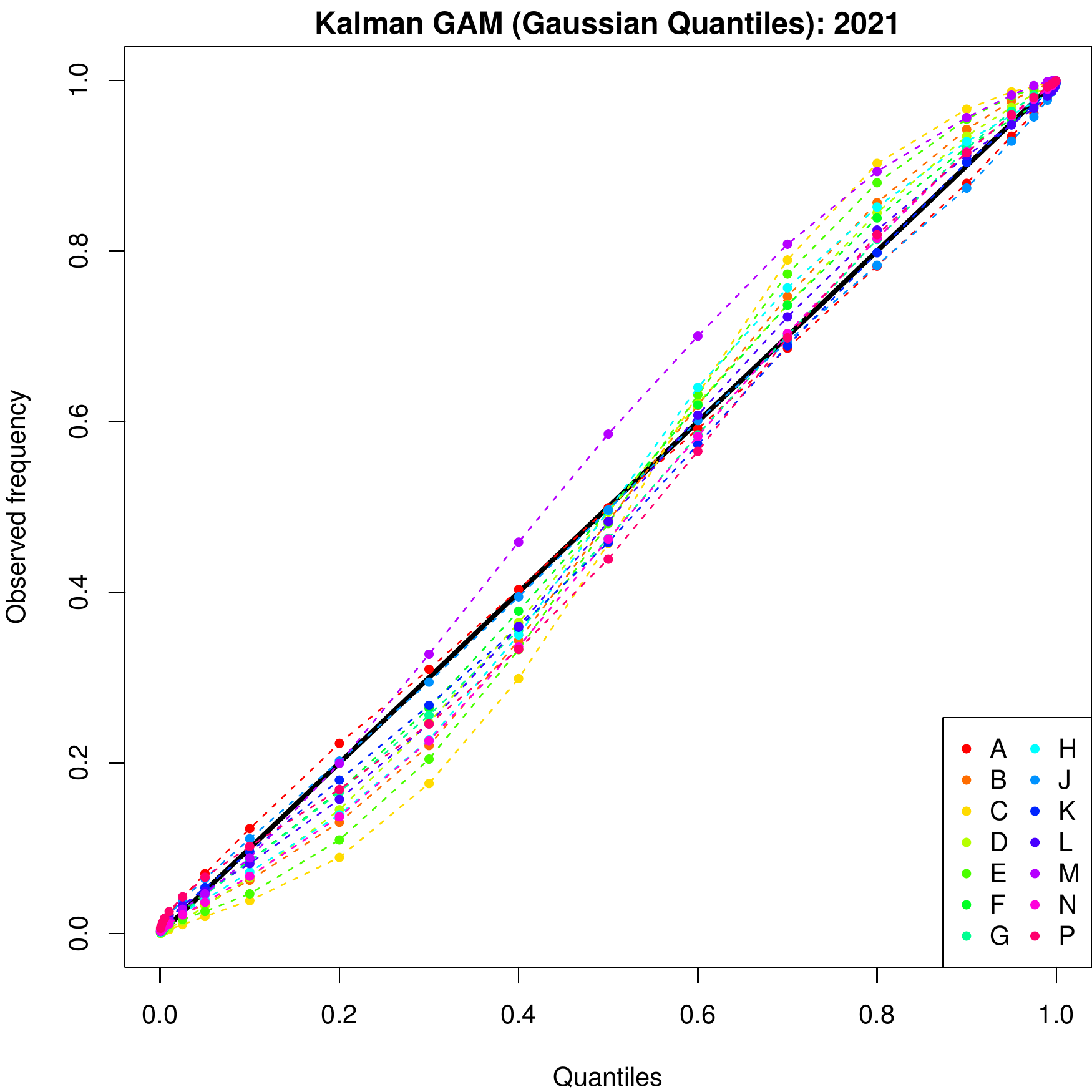}
\hspace{0.5cm}
\includegraphics[width=5.4cm]{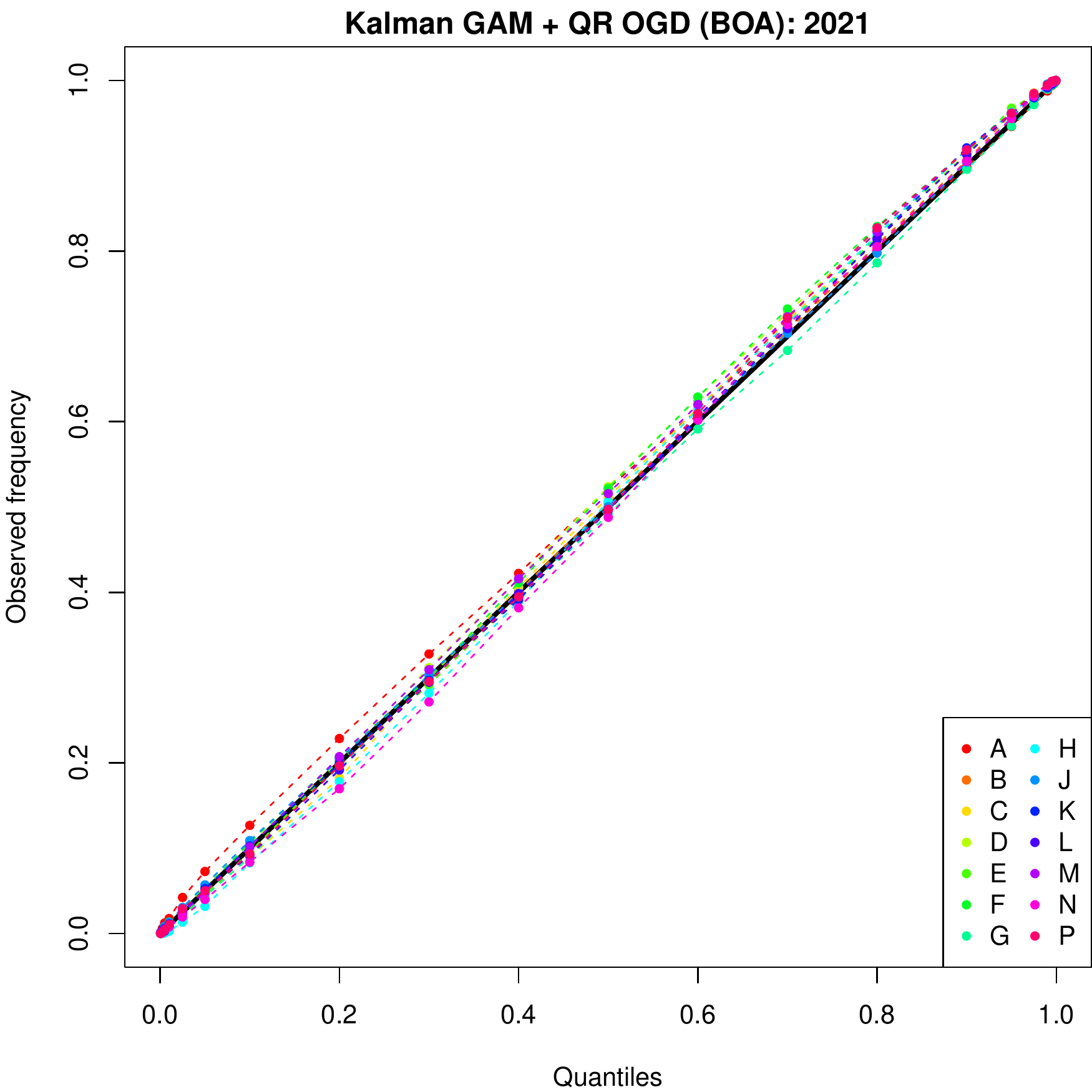}
\caption{Reliability diagrams for different methods in 2019 and in 2021.}
\label{fig:reliabilityGB}
\end{figure*}

\begin{figure*}
    \centering
    \includegraphics[width=5.8cm]{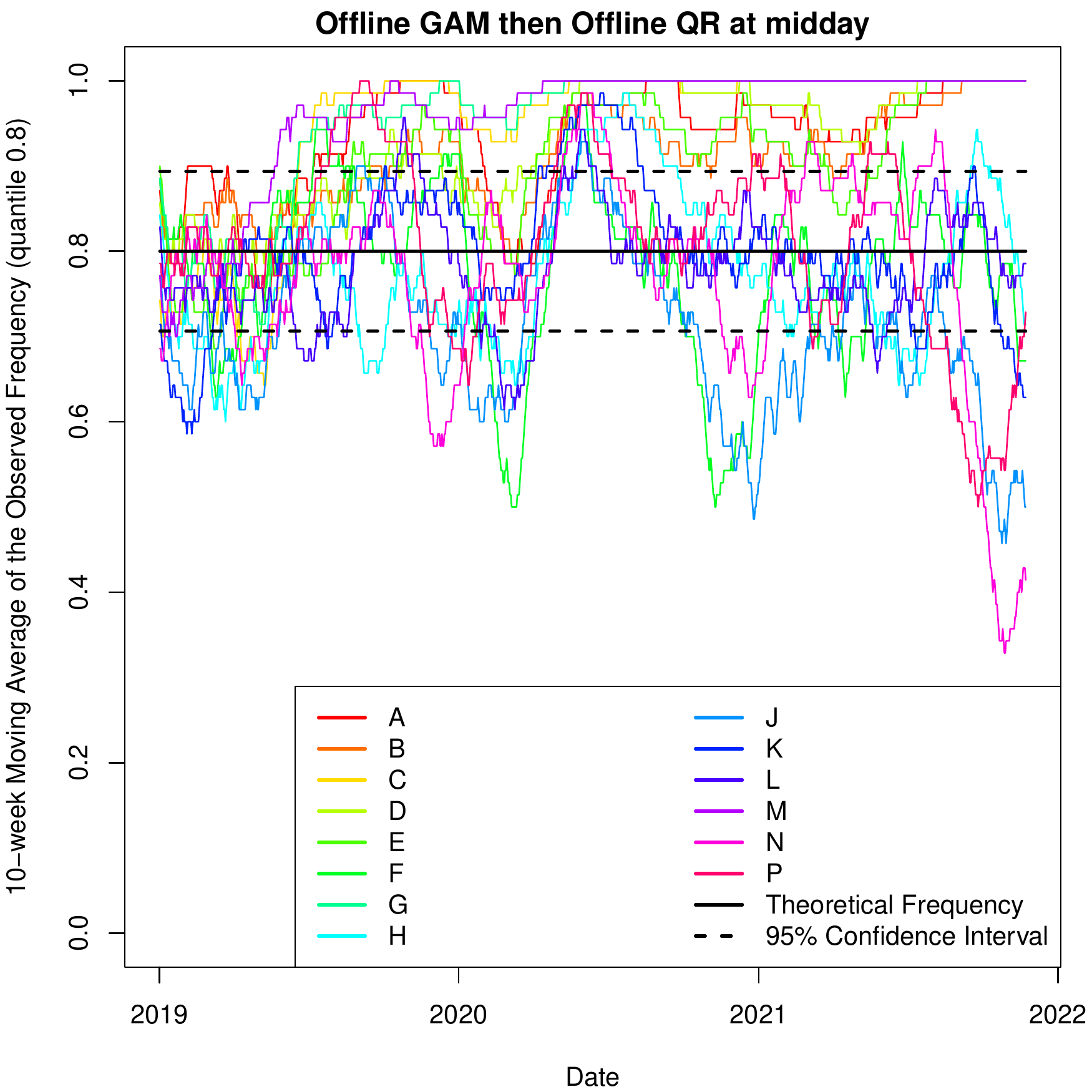}
    \hspace{0.5cm}
    \includegraphics[width=5.8cm]{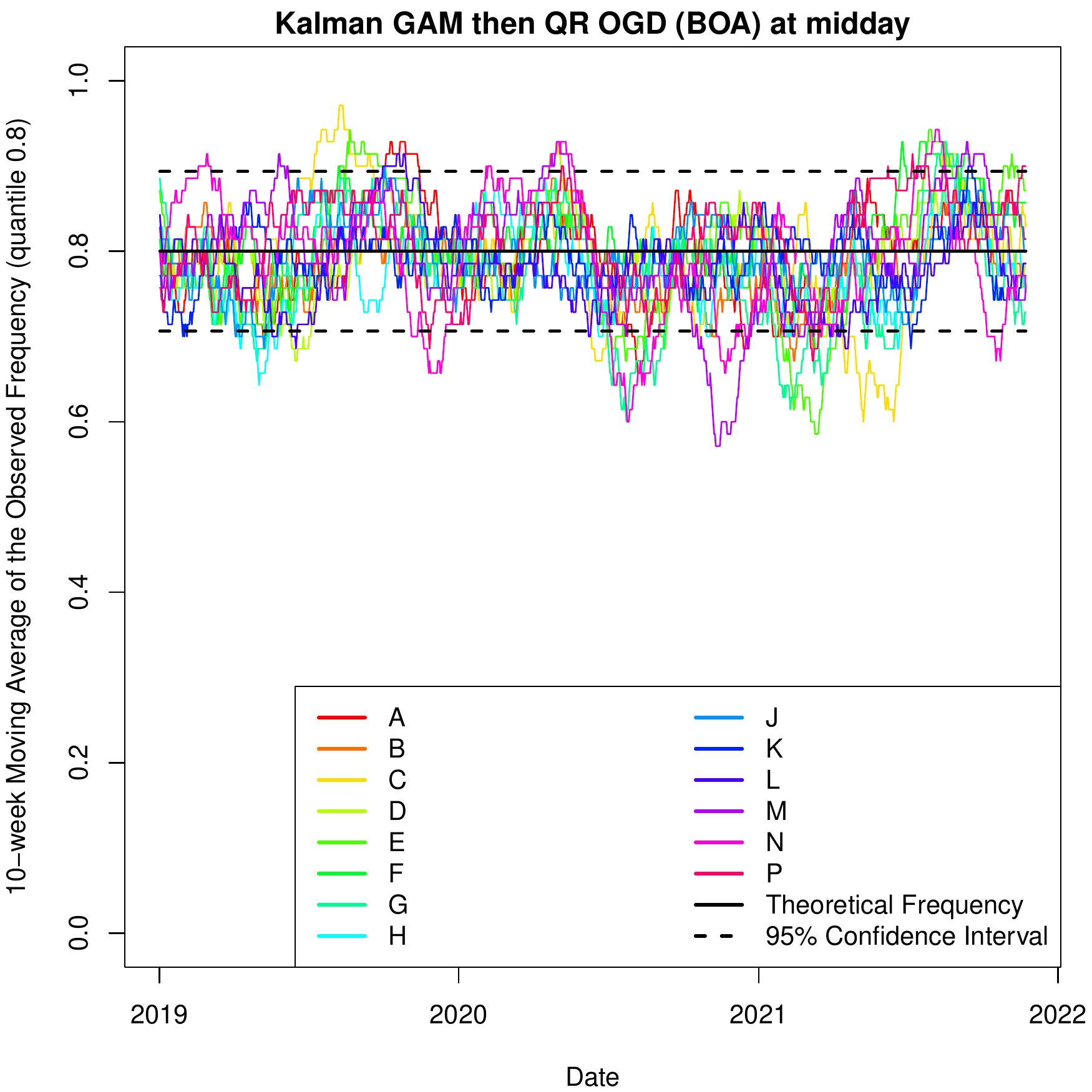}
    \caption{Evolution of the calibration. We select a specific time of day (midday) and a quantile level 0.8; we display the observed frequency that the net-load exceeds the quantile forecast. The 95\% confidence interval is based on the assumption that the residuals are iid. We see that for the offline model (GAM then quantile regression) the confidence interval is not satisfied.}
    \label{fig:GBerror8evol}
\end{figure*}

We display reliability diagrams in Figure~\ref{fig:reliabilityGB}. We observe that the offline model is not reliable, probably because of the biases in the mean model.
The adaptation of the mean model by the Kalman filter yields Gaussian quantiles that are much more reliable.
Reliability is then improved further by the adaptation of the quantile regressions.
Similarly to the evolution of the bias of the model (Figure~\ref{fig:GBerrorevol}), we display in Figure~\ref{fig:GBerror8evol} the evolution of the observed frequency for a specific quantile. Our adaptive model has a much better evolution of calibration in time than the offline version.

We provide in Table~\ref{tab:rps} the RPS averaged on the 14 regions for the different models.
\begin{table}
\caption{Aggregate nRPS on the 14 regions for the different models.}
\label{tab:rps}
\centering
\begin{tabular}{c|c|c|c}
	& 2019 & 2020 & 2021\\
	\hline
	Offline Method & 0.231 & 0.338 & 0.454 \\
	\hline
	GAM Kalman (Gaussian Quantiles) & 0.212 & 0.217 & 0.222 \\
	\hline
	GAM Kalman + Offline QR & {\bf 0.206} & {\bf 0.214} & {\bf 0.217} \\
	\hline
	Offline GAM + QR OGD ($10^{-3}$) & 0.218 & 0.270 & 0.293 \\
	Offline GAM + QR OGD ($10^{-2}$) & 0.207 & 0.221 & 0.218 \\
	Offline GAM + QR OGD ($10^{-1}$) & 0.250 & 0.248 & 0.293 \\
	\hline
	Offline GAM + QR OGD (BOA) & 0.204 & 0.211 & 0.216 \\
	\hline
	GAM Kalman + QR OGD ($10^{-2}$) & 0.205 & 0.204 & 0.212 \\
	\hline
	GAM Kalman + QR OGD (BOA) & {\bf 0.202} & {\bf 0.201} & {\bf 0.209}
\end{tabular}
\end{table}
We obtain an important gain in nRPS by adapting the GAM using the Kalman filter and keeping an offline quantile regression, even for the stable period (the year 2019).
Adapting the quantile regression with an OGD has a very tenuous difference. The step size of this OGD may be selected by Bernstein Online Aggregation, combining step sizes in $\{10^i, -8\le i\le 0\}$. This choice of step size does not only perform as well as the best expert as suggested in Section~\ref{sec:aggregation}, but it outperforms this oracle. Combining both levels of adaptation yields the best performances.
\begin{figure}
    \centering
    \includegraphics[width=6cm]{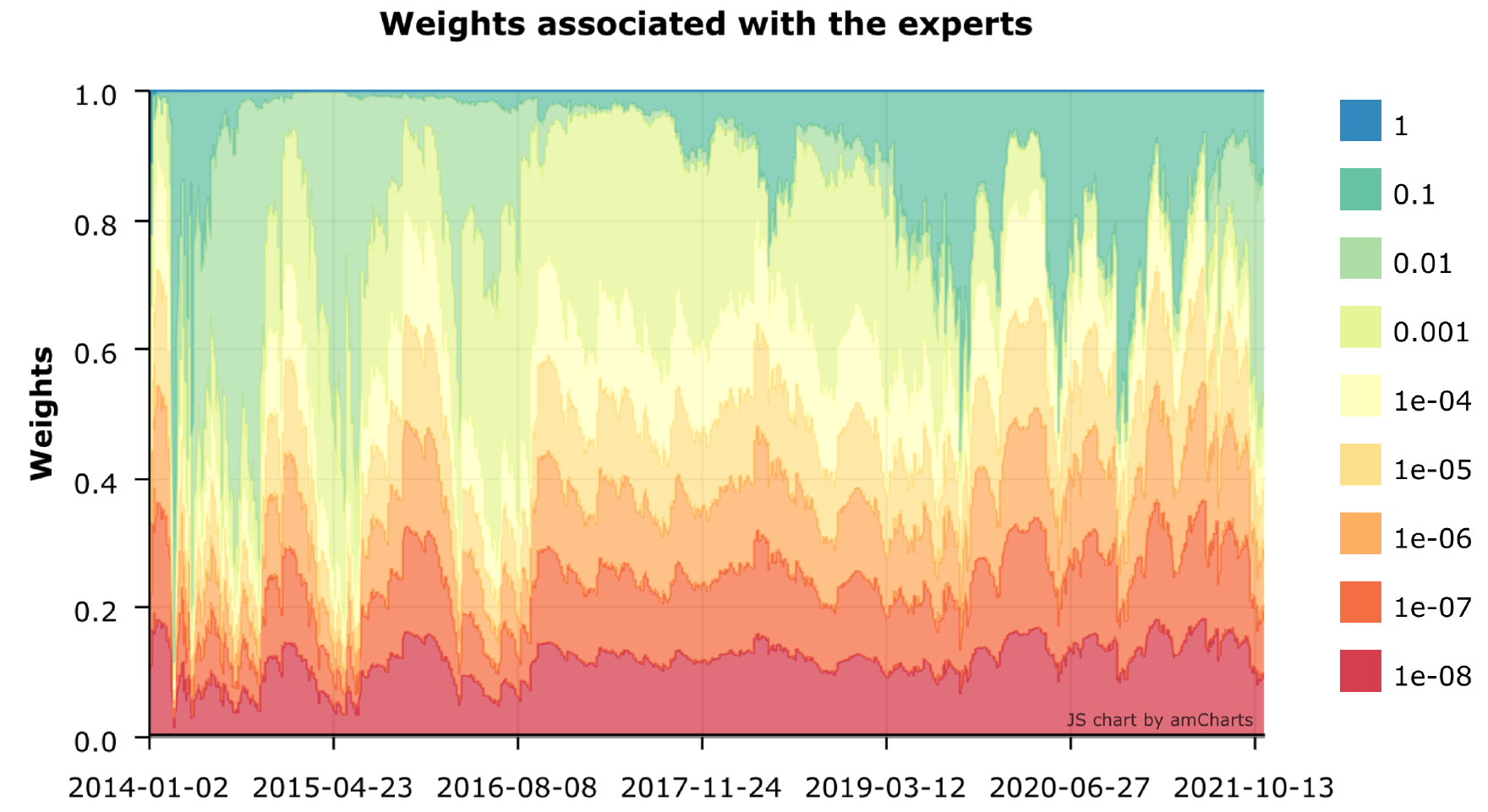}
    \caption{Weights optimized by BOA for the forecast of region A in Great Britain at quantile level 0.1.}
    \label{fig:boa}
\end{figure}

\subsection{Learning the Embedded Capacities}\label{sec:rmcap}
A remarkable advantage of model adaptation is that it reduces the need for good explanatory variables, which may be difficult to obtain. In particular, we show that our Kalman GAM can {\it learn} the embedded generation capacities. That is removing these variables does not significantly change the predictions, contrary to the offline method.

More precisely, in the GAM and in the quantile regressions presented in Section~\ref{sec:GBoffline}, we remove the solar and wind generation capacities. We consider solar radiation only instead of its product with solar capacity. In the nonlinear GAM effect of the wind speed, we remove the dependence on the embedded wind capacity. We thus evaluate removing the capacities in the offline model, as well as in the Kalman adaptation of the GAM. The evolution of the capacities may be captured by the adaptation of the state coefficient.

We evaluate during 2019 to not include the coronavirus crisis.
For mean prediction, removing the capacities increases the nRMSE of the offline method by more than $10\%$, while it reduces the nRMSE by $0.4\%$ for the Kalman adaptation of the GAM.
For probabilistic prediction, removing the capacities increases the nRPS by more than $10\%$ for the offline model and reduces it by $0.02\%$ for the offline quantile regression on the residuals of GAM Kalman.

\section{City-Wide Load in the United States}\label{sec:US}
We test our framework also on US data during the coronavirus crisis \cite{ruan2020cross}. To demonstrate the generality of the approach, we consider a very different setting. We forecast the standard load at a daily granularity and with no delay in data availability. Each day, we can adapt on the data up to the current day in order to forecast the next.

\subsection{Data Presentation and Offline Model}
\begin{figure}
\centering
\includegraphics[width=4.3cm]{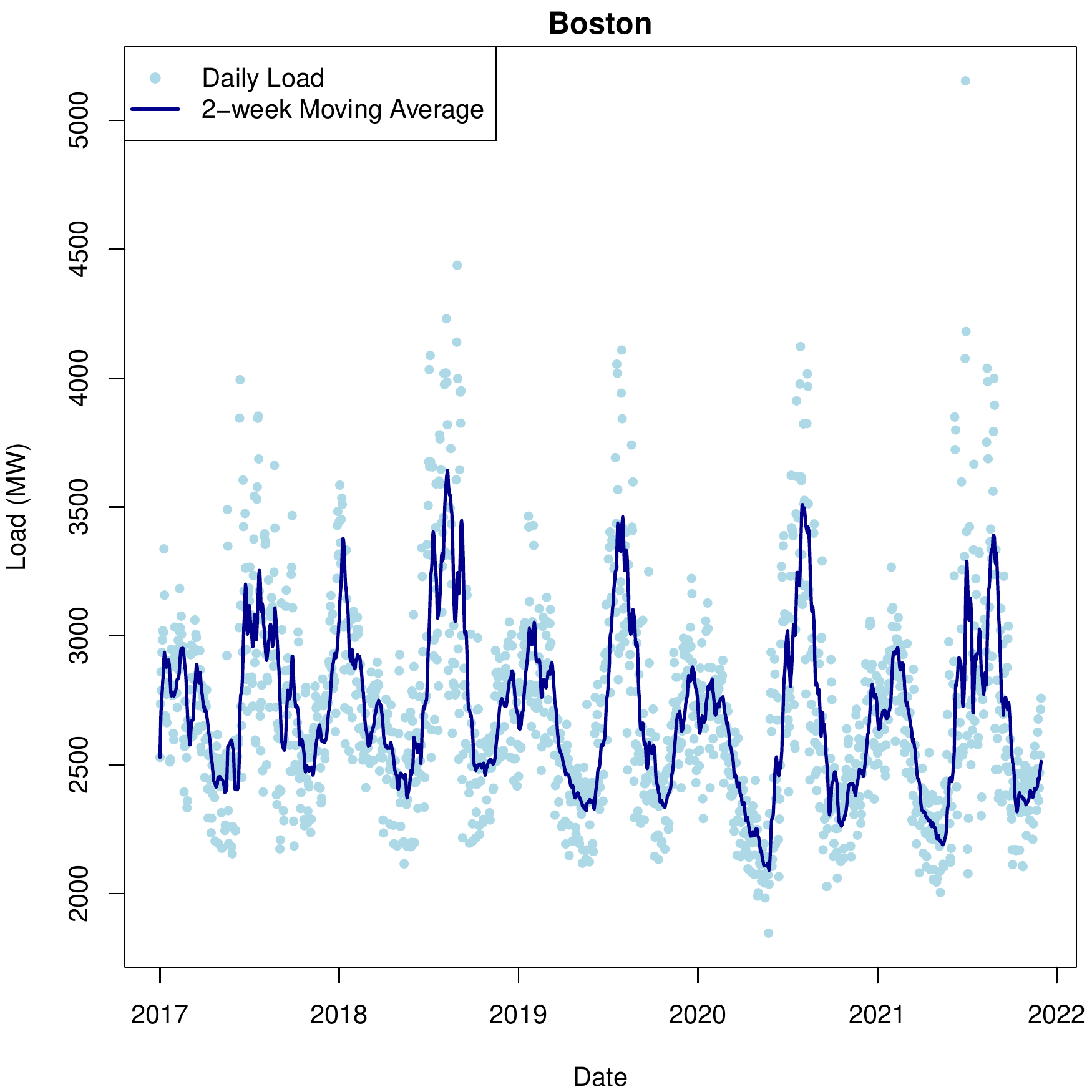}
\includegraphics[width=4.3cm]{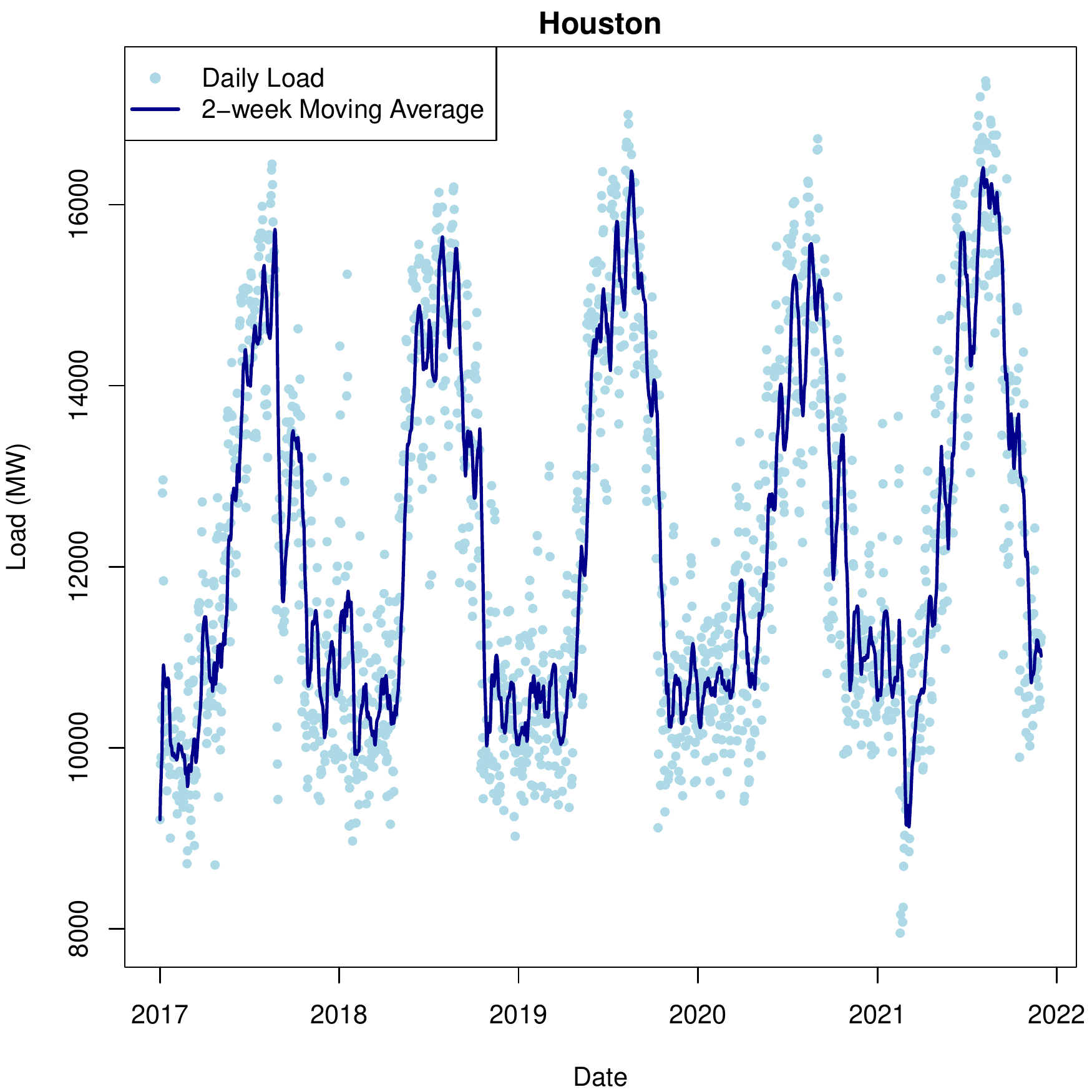}
\caption{Daily electricity load in Boston (left) and Houston (right). The consumption has two annual peaks in summer and winter in Boston, whereas there is only a summer peak in Houston.}
\label{fig:USload}
\end{figure}

The data set covers seven U.S. electricity markets, and the major city for each one: Boston, Chicago, Houston, Kansas City, Los Angeles, New York City, Philadelphia.
Consider the city-wide loads shown in Figure~\ref{fig:USload} and note that load patterns are different across cities.

We follow the forecasting structure presented in Section~\ref{sec:theory}.
\begin{enumerate}
\item
The generalized additive model we consider has the following form:
\begin{align}
	\nonumber
 Load_t & = \sum\limits_{i=1}^7 \alpha_i \mathds{1}_{WeekDay_t=i} + \beta_1 BH_t + \beta_2 WB_t \\
	& \quad + \beta_3 LoadD_t + f_1(LoadW_t) + f_2(t) \\
	\nonumber
 & \quad + f_3(Temp_t) + f_4(Hum_t) + f_5(Toy_t) + \varepsilon_t \,,
\end{align}
where $\varepsilon_t$ is a Gaussian i.i.d. noise and for each day $t$,
\begin{itemize}
\item
$WeekDay_t$ is the day of the week,
\item
$BH_t$ (respectively $WB_t$) is a Boolean denoting if the day is a bank holiday (respectively in the winter break),
\item
$LoadD_t$ and $LoadW_t$ are lags of the load with a one-day and one-week delays,
\item
$t$ is the day (variable growing linearly with time),
\item
$Temp_t$ and $Hum_t$ are the temperature and humidity,
\item
$Toy_t$ is the time of year (variable growing linearly from 0 on January 1\textsuperscript{st} to 1 on December 31\textsuperscript{st}).
\end{itemize}
The nonlinear effects $f_1,f_2,f_3,f_4,f_5$ are built using thin-plate splines for the first four, and cubic cyclic splines for $f_5$.

\item
The quantile regression we fit on the residuals uses as covariates the effects of the GAM, the mean prediction, and the square of the mean prediction as in Section \ref{sec:GBoffline}:
\end{enumerate}
\begin{align}
    \nonumber
    z_t & = \Big(\hat y_t, \hat y_t^2, \sum\limits_{i=1}^7 \alpha_i \mathds{1}_{WeekDay_t=i}, \beta_1 BH_t, \beta_2 WB_t, \\
    & \qquad \beta_3 LoadD_t, f_1(LoadW_t), f_2(t), f_3(Temp_t),\\ 
    \nonumber
    & \qquad f_4(Hum_t), f_5(Toy_t), 1\Big)^\top \,.
\end{align}

\subsection{Evaluation}
We display the nRMSE and nMAE aggregated on the seven cities in Table~\ref{tab:USerror}.
\begin{table}
    \caption{Aggregate metrics in the seven US cities.}
    \label{tab:USerror}
    \centering
    \begin{tabular}{c|c c|c c}
        & \multicolumn{2}{c|}{2020} & \multicolumn{2}{c}{2021} \\
        \hline
        Forecast & nRMSE & nMAE & nRMSE & nMAE \\
        \hline
        Persistence (7 days) & 0.777 & 0.688 & 0.852 & 0.745 \\ 
        Persistence (1 day) & 0.455 & 0.417 & 0.464 & 0.414 \\ 
        \hline 
        GAM & 0.407 & 0.400 & 0.666 & 0.616 \\
        Incremental GAM (daily) & 0.223 & 0.213 & 0.214 & 0.185 \\
        GAM KF Static & 0.206 & 0.195 & 0.204 & 0.178  \\
        GAM KF Dynamic & {\bf 0.194} & {\bf 0.168} & {\bf 0.198} & {\bf 0.166}
    \end{tabular}
\end{table}
The nRPS is displayed in Table~\ref{tab:USrps}.
\begin{table}[]
    \caption{Aggregate nRPS in the seven US cities.}
    \label{tab:USrps}
    \centering
    \begin{tabular}{c|c|c}
        Forecast & 2020 & 2021 \\
        \hline
        Offline method & 0.319 & 0.539 \\
        \hline
        GAM Kalman (Gaussian Quantiles) & 0.122 & 0.123 \\
        \hline
        GAM Kalman + Offline QR & 0.230 & 0.391 \\
        \hline
        Offline GAM + QR OGD ($10^{-4}$) & 0.232 & 0.157 \\
        Offline GAM + QR OGD ($10^{-3}$) & 0.145 & 0.109 \\
        Offline GAM + QR OGD ($10^{-2}$) & 0.109 & 0.196 \\
        Offline GAM + QR OGD ($10^{-1}$) & 0.612 & 1.75 \\
        \hline
        Offline GAM + QR OGD (BOA) & 0.099 & 0.097 \\
        \hline
        GAM Kalman + QR OGD (BOA) & {\bf 0.094} & {\bf 0.094}
    \end{tabular}
\end{table}

The best models are still the dynamic setting of the Kalman filter for mean forecasting and the combination of Kalman adaptation of GAM with the adaptive quantile regression for the probabilistic task. However, the performances are significantly different from the results of Section~\ref{sec:GB}. In particular, adapting the mean model with Kalman and then applying offline quantile regression models behaves poorly. It is crucial to adapt the quantile models.

\section{Conclusion}
In this article, we applied an adaptive procedure for probabilistic net-load forecasting.
The proposed methodology relies on several steps: an adaptive mean forecast is obtained by the Kalman filter, and adaptive quantile regressions are derived from Online Gradient Descent. To solve the choice of the step size in the OGD, the algorithm is estimated with multiple learning rates and the estimates are combined using Bernstein Online Aggregation.

Evaluation has been performed based on the electricity net-load in the fourteen regions of Great Britain, and on electricity demand in seven major cities in the United States.
Online adaptation of the mean model is always important, yielding reductions in RMSE of around 10\% in all case studies compared to updating the model periodically. Additional adaptation of quantile regressions added marginal benefits for regional load forecasting in GB, but substantial benefits for US cities, reducing the RPS by over 20\%. This may be due to the fact that the coronavirus crisis has had more impact on daily lives in big cities. Therefore, in our applications, adaptive methods have more interest for big cities. Indeed, we also observe a more substantial gain for region C (London) in the GB data set. 

A natural extension would be the adaptation of multivariate probabilistic forecasts. Previous works have considered adaptive dependency structures, but how models for marginal distributions and dependency structures could (and should) be adapted simultaneously has not been studied.
Also, we focused on non-extreme quantiles and adaptive estimation of extreme values is an exciting and very challenging topic. A special treatment for extremes may be useful to refine reliability as well as numerical performance.


\bibliographystyle{IEEEtran}
\bibliography{IEEEabrv,mybib.bib}

\vfill

\end{document}